\documentclass[preprintnumbers, twocolumn, preprintnumbers, PRD, superscriptaddress, nofootinbib]{revtex4-2}
\usepackage{hyperref}
\hypersetup{
  colorlinks=true,        
  linkcolor=blue,         
  citecolor=cyan,         
}
\usepackage{cancel}
\usepackage{amsfonts}
\usepackage{amsmath}
\usepackage{amssymb,epsf}
\usepackage{latexsym}
\usepackage{graphicx,epsfig}
\usepackage{amssymb}
\usepackage{float}
\usepackage{epstopdf}
\usepackage{dcolumn}
\usepackage{psfrag}
\usepackage{wrapfig}
\usepackage{makeidx}
\usepackage{epsf}
\usepackage{color}
\usepackage{multirow}
\usepackage{tabularx}
\usepackage{array}
\numberwithin{equation}{section}
\def\be {\begin{equation}}
\def\ee {\end{equation}}
\def\bea {\begin{eqnarray}}
\def\eea {\end{eqnarray}}
\def\bc {\begin{center}}
\def\ec {\end{center}}
\def\bfg {\begin{figure}}
\def\efg {\end{figure}}
\def\bi {\begin{itemize}}
\def\ei {\end{itemize}}

\def\beq{\begin{equation}}
\def\eeq{\end{equation}}
\def\br{\begin{eqnarray}}
\def\er{\end{eqnarray}}
\newcommand{\eel}[1] {\label{#1}\end{equation}}

\begin{document}
\title{\Large Cloud Screening of extremal charged BTZ black hole}

\author{Mendrit Latifi}
\email{ml67500@student.uni-lj.si} 
\affiliation{Faculty of Mathematics and Physics, University of Ljubljana,
Jadranska ulica 19, SI-1000 Ljubljana, Slovenia}

\begin{abstract}
We study the instability of a charged scalar field in the near-horizon region of an extremal charged BTZ black hole. The extremal geometry contains an $\mathrm{AdS}_2\times S^1$ throat supported by a constant electric field, which lowers the effective scalar mass and can violate the $\mathrm{AdS}_2$ Breitenlohner--Freedman bound. This identifies an infrared instability of the extremal normal state. We then construct a nearby finite-temperature branch of fully backreacted, source-free, node-free hairy black-hole solutions, thereby establishing the nonlinear continuation of the unstable mode near its onset. The scalar condensate modifies both the Maxwell field and the geometry, while stronger gravitational backreaction lowers the critical temperature and suppresses the condensate amplitude. Near the critical point, the condensate exhibits mean-field scaling. We also develop an effective near-horizon interpretation in terms of electric-flux screening. The exact zero-temperature endpoint of the hairy branch is not determined here and remains an open problem.
\end{abstract}
\maketitle

\section{\label{sec:level1}Introduction}
Black holes provide a natural setting for studying the interplay between gravity, quantum field theory, and strong-field dynamics. Although early no-hair theorems suggested that stationary black holes are characterized only by a small set of global charges
\cite{Israel:1967za,Ruffini:1971bza}, these results rely on restrictive assumptions. Once these assumptions are relaxed, rotating and charged black holes may support nontrivial scalar or vector configurations, including linear clouds at the threshold of instability and fully nonlinear hairy solutions
\cite{hod2013no,hod2015extremal,santos2020black,Herdeiro:2016tmi}.\newline
In asymptotically AdS spacetimes, boundary conditions play a central role in determining which field configurations are physically admissible. In three dimensions, this issue is particularly subtle for electrically charged black holes. For nonvanishing electric charge, the original rotating BTZ ansatz solves the AdS Einstein--Maxwell equations only in the non-rotating case
\cite{clement1993classical,carlip19952+,Carlip:1995qv,kamata1995electrically,Cataldo:1999fh,Garcia:1999py}. A consistent charged rotating solution was subsequently constructed by Mart\'{\i}nez, Teitelboim, and Zanelli \cite{Martinez:1999qi}, with the static charged BTZ black hole obtained in the $J\to0$ limit. Scalar fields in BTZ backgrounds may also require mixed, or Robin, boundary conditions in order to define a well-posed variational principle and support nontrivial profiles \cite{Ferreira:2017cta,dappiaggi2018superradiance}.\newline
In this work, we study scalar condensation driven by the electric field rather than by rotation. We focus on extremal charged BTZ black holes, whose near-horizon geometry contains an $\mathrm{AdS}_2\times S^1$ throat supported by a constant electric field
\cite{Cadoni:2008mw,azeyanagi2008near}. A charged scalar in this throat experiences an electric-field-dependent reduction of its effective $\mathrm{AdS}_2$ mass. Above a critical electric field, the effective mass violates the near-horizon Breitenlohner--Freedman bound
\cite{Breitenlohner:1982bm,Breitenlohner:1982jf}, signaling an infrared instability. This instability may be interpreted as charged pair production in the near-horizon electric field, followed by the formation of a scalar condensate that redistributes and partially screens the electric flux.\newline
The relation between charged-scalar condensation and violation of the near-horizon $\mathrm{AdS}_2$ BF bound was emphasized in the study of charged AdS black holes
\cite{Gubser:2008px} and became central to holographic superconductor constructions
\cite{Hartnoll:2008vx,Hartnoll:2008kx}. In these models, the unstable scalar mode continues into a backreacted hairy black-hole phase. Related backreacted superconducting solutions in BTZ backgrounds were studied in
\cite{Liu:2011fy}, while the broader role of near-horizon BF bounds in the stability of AdS black holes was investigated in
\cite{Dias:2010ma,Basu:2010uz,Dias:2011tj}.\newline
It is important to distinguish this mechanism from instabilities induced primarily by mixed or double-trace boundary conditions
\cite{Faulkner:2010gj,Allais:2010qq,Basu:2013soa,Dias:2013bwa,Ren:2022qkr,Kinoshita:2023iad,Auzzi:2025nxx}. Such boundary-driven instabilities may occur even in pure AdS when the boundary coupling lies in an appropriate range
\cite{Ishibashi:2004wx}, and have recently been studied for static and rotating BTZ black holes
\cite{Dias:2025uyk}. By contrast, the instability considered here is triggered by the electric field and the violation of the near-horizon $\mathrm{AdS}_2$ BF bound. Mixed boundary conditions remain relevant for the variational principle and holographic interpretation, but they are not the primary origin of the instability.\newline
The near-horizon description also connects the problem to infrared criticality in holographic systems. Charged black holes with an $\mathrm{AdS}_2$ throat are governed at low energies by an effective $\mathrm{CFT}_1$
\cite{iqbal2010quantum,faulkner2011emergent,Iqbal:2011aj}. When the effective $\mathrm{AdS}_2$ mass violates the BF bound, the infrared scaling dimension becomes complex, indicating an instability of the infrared fixed point. Related behavior appears in studies of charged-black-hole response functions, quasinormal modes, and late-time dynamics
\cite{Hod:2012px,Degollado:2014vsa,fontana2024quasinormal}.\newline
A complementary interpretation is provided by Wilson lines and defect conformal field theories, where localized degrees of freedom can generate screening clouds around defects
\cite{Aharony:2022ntz,Aharony:2023amq}. From this perspective, the charged BTZ black hole acts as a dynamical impurity, while the scalar condensate represents its infrared screening response.\newline
This paper is organized as follows. In Sec.~\ref{sec:level2}, we review the charged BTZ background, its extremal limit, and the emergence of the near-horizon $\mathrm{AdS}_2\times S^1$ geometry. In Sec.~\ref{sec:level3}, we analyze the charged scalar and derive the condition for violation of the effective $\mathrm{AdS}_2$ BF bound. In Sec.~\ref{sec:level4}, we include the gravitational and Maxwell backreaction and construct the nonlinear hairy branch. The subsequent sections discuss the boundary conditions, the effective screening description, and the physical interpretation of the solutions. We conclude in Sec.~\ref{sec:level6}.


\section{\label{sec:level2}Background Geometry}
We start with the following action \cite{Cadoni:2008mw}:
\begin{equation} \label{EM-action}
    I = \frac{1}{16 \pi G} \int d^3x \sqrt{-g} \left[ R+ \frac{2}{\ell^2} -4 \pi G F_{\mu \nu} F^{\mu \nu} \right].
\end{equation}
A solution to a static charged BTZ black hole is given by the following metric and gauge field, respectively:
\begin{equation} \label{fullmetric}
    ds_3^2= -f(r)dt^2 + f^{-1}(r)dr^2 +r^2 d\theta^2 
\end{equation}
where $f(r)$ and $F_{tr}$ are given by:
\begin{align} \label{lapsefunc}
    &f(r)= -8G M +\frac{r^2}{\ell^2} - 8 \pi G Q^2 \ln{\left( \frac{r}{\ell}\right)}, \quad \text{and} \\
    & \quad F_{tr} = \frac{Q}{r}.
\end{align}
where by setting $f(r)=0$ we find zeros of $f(r)$, namely $r_{\pm}$, where 
\begin{equation}
    \Delta= 8GM-4 \pi G Q^2 \left[ 1-2 \ln{\left( 2Q \sqrt{\pi G} \right)} \right].
\end{equation}
For 
\begin{equation}
    \Delta = \begin{cases}
        >0, \text{two horizons} \\
        =0, \text{extremal(double root)} \\
        <0, \text{none}.
    \end{cases}
\end{equation}
The temperature is then given by \cite{Hawking:1975vcx,GibbonsHawking}
\begin{equation}
    T_h= \frac{r_+}{2 \pi \ell^2} - \frac{2G Q^2}{r_+}
\end{equation}
and the entropy \cite{Hawking:1975vcx,Bekenstein}
\begin{equation}
    S= \frac{\text{perimeter}}{4G} = \frac{2 \pi r_+}{4G} =\frac{\pi r_+}{2G}. 
\end{equation}
Extremality is then found by setting 
\begin{equation}
    f(\gamma)= 0 \quad \text{and} \quad f^{\prime}(\gamma)=0. 
\end{equation}
From $f^{\prime}(\gamma)=0$ we get
\begin{equation} \label{circleradius}
    \gamma^2=4 \pi GQ^2\ell^2.
\end{equation}
From $f(\gamma)=0$ we find the extremal mass $M_{\rm ext}$
\begin{equation} \label{extremalmass}
    M_{\rm ext} = M(\gamma)= \pi Q^2 \left[\frac{1}{2} - \ln{\left( 2Q \sqrt{\pi G}\right)} \right].
\end{equation}
At extremality we have $T_H=0$ and $S_{\rm ext}= \frac{\pi \gamma}{2G} \neq 0$. Extremal BTZ is a \emph{zero-temp}, finite entropy "ground state". \newline
Shifting now to $r=\gamma+x$ for $\mid x\mid \ll \gamma$, we can write the lapse function as \cite{Cadoni:2008mw}
\begin{equation}
    f(\gamma +x) =\frac{2}{\ell^2} x^2- 8G \Delta M + \mathcal{O}(x^3),
\end{equation}
which is the lapse function for near-horizon, near-extremal limit and 
\begin{equation}
    F_{tx}=\frac{1}{2\sqrt{\pi G}\ell} + \mathcal{O} (x).
\end{equation}
where at horizon, $F_{tx}$ becomes a constant, i.e  the throat sees a uniform electric field. \newline 
The near-horizon, near-extremal metric of the charged BTZ black hole will be 
\begin{equation}\label{near-extremal.metric}    
\begin{split}
    ds^2 =& - \left(\frac{2}{\ell^2} x^2- 8G \Delta M  \right)dt^2 \\
    & +\left(\frac{2}{\ell^2} x^2- 8G \Delta M  \right)^{-1}dx^2  \\
     & +\gamma^2 d\theta^2
\end{split}
\end{equation}
and
\begin{equation}
    F_{tx}=\frac{1}{2\sqrt{\pi G}\ell} 
\end{equation}
where
\begin{equation}
\begin{split}
    \Delta M &=M -M(\gamma) \\
   &  = M - \pi Q^2 \left[\frac{1}{2} - \ln{\left( 2Q \sqrt{\pi G}\right)} \right].
\end{split}
\end{equation}
where $\Delta M$ is the mass above extremality. 
As for now we will only work with \emph{extremal metric}. Taking $\Delta M\to 0$ we find the extremal metric to be: 
\begin{equation}\label{extremal.metric}    
\begin{split} 
    ds^2 =& - \left(\frac{2}{\ell^2} x^2 \right)dt^2 \\
    & +\left(\frac{2}{\ell^2} x^2  \right)^{-1}dx^2  \\
     & +\gamma^2 d\theta^2.
\end{split}
\end{equation}
The $(t,x)$-part is the $\rm AdS_2$  with radius $R_A^2= \frac{\ell^2}{2}$ and the circle $S^1$ has a fixed radius $\gamma = 2\sqrt{\pi G}  \mid Q \mid  \ell$. Under dimensional reduction to $\rm 2D$, this fixed radius is a constant dilaton.\newline
Finally, introducing the Poincare coordinate 
\begin{equation} \label{Poincarecoordinate}
    z= \frac{R_A^2}{x}
\end{equation}
the two-dimensional part of the metric takes the standard form 
\begin{equation}
    ds_{\rm AdS_2}^2 = \frac{R_A^2}{z^2} \left( -dt^2 +dz^2  \right)
\end{equation}
where the $\rm AdS_2$ boundary is at $z \to 0$ while the horizon is at $z \to  \infty$. \newline
The above geometry is the exact scalar-free solution of the Einstein-Maxwell equations. In the following sections it will be used as the zeroth-order background for the linear charged-scalar instability. This distinction is important. At the onset of the instability, the charged scalar may consistently be treated as a linear perturbation on the fixed charged-BTZ background. A finite scalar condensate, however, sources both the Maxwell equation and Einstein's equations through its current and stress tensor. Therefore a finite-amplitude screened scalar-cloud configuration requires solving the fully backreacted Einstein--Maxwell--scalar system. The fixed $\mathrm{AdS}_2\times S^1$ throat derived above is thus the correct background for the linear onset problem and for the probe-limit analysis, while the backreacted condensate branch will be discussed separately below.\newline
The extremal throat has three important features. First, extremality gives a double zero of the lapse function, producing the quadratic near-horizon potential $f(r)\sim x^2$. Second, the circle remains of fixed radius, $\gamma=\mathrm{const.}$, so dimensional reduction gives a constant-dilaton $\mathrm{AdS}_2$ geometry. Third, the electric field approaches a constant at the horizon, providing a uniform background electric field in the effective $\mathrm{AdS}_2$ description.\newline
Scalar clouds and black-hole hair have been studied in many related settings. In four dimensions, stationary scalar configurations around Kerr-Newman and Reissner-Nordström black holes, as well as boson-star-like configurations, often arise through superradiant amplification or through a balance between gravitational, electromagnetic, and boundary-condition effects \cite{Hod:2012px,hod2015extremal,Herdeiro:2014goa}. More broadly, these examples fit into the long-standing discussion of black-hole no-hair theorems and their possible loopholes \cite{Israel:1967za,Ruffini:1971bza,Volkov:1998cc,Volkov:2016ehx,Herdeiro:2016tmi}. Related dynamical signatures of massive fields around black holes, including long-lived tails and possible gravitational-wave imprints, have also been discussed in the literature \cite{Degollado:2014vsa}. In three-dimensional gravity, scalar hair and boson-star--like solutions are known in asymptotically $\mathrm{AdS}_3$ spacetimes, often supported by self-interactions or by nontrivial boundary conditions \cite{Henneaux:2002wm,Martinez:2004nb}.\newline
The mechanism studied here is different. The scalar cloud is driven by the infrared physics of the extremal charged-BTZ throat: the electric field lowers the effective mass of the charged scalar in the near-horizon $\mathrm{AdS}_2$ region, leading to a BF-bound instability that may be interpreted as a Schwinger-type screening instability. Thus the cloud arises from local near-horizon dynamics rather than from global superradiant amplification, rotation, or scalar self-interactions. This provides a controlled setting in which hair formation is governed by the $\mathrm{AdS}_2$ region while remaining embedded in the full three-dimensional charged-BTZ geometry.

\section{\label{sec:level3}Klein-Gordon equation}
We now study the onset of the charged-scalar instability on the extremal charged-BTZ background derived in Sec.\ref{sec:level2}. In this section the metric and gauge field are held fixed at their zeroth-order values, while the scalar is treated as a linear perturbation, 
\begin{equation}
    g_{\mu \nu}= g_{\mu \nu}^{(0)}, \quad A_{\mu}= A_{\mu}^{(0)}, \quad \Phi= \mathcal{O}(\epsilon).
\end{equation}
Backreaction of the scalar on the Maxwell field and on the geometry begins at quadratic order in the scalar amplitude and will be discussed separately. \newline
We consider a minimally coupled, charged scalar field $\Phi$ with mass $m$ and charge $q$ described by the action
\begin{equation}
S = \int d^3 x \sqrt{-g} \left( -\frac{1}{2} D_\mu \Phi^* D^\mu \Phi - \frac{1}{2} m^2 \Phi^* \Phi \right),
\end{equation}
where $D_\mu = \nabla_\mu - i q A_\mu$ is the gauge-covariant derivative. The field $\Phi$ satisfies the Klein–Gordon equation:
\begin{equation} \label{kleingordon}
\left( \nabla_\mu - i q A_\mu \right) \left( \nabla^\mu - i q A^\mu \right) \Phi - m^2 \Phi = 0.
\end{equation}
which can also be written as 
\begin{equation}
    \frac{1}{\sqrt{-g}} D_{\mu} \left( \sqrt{-g} g^{\mu \nu} D_\nu \Phi \right) -m^2\Phi=0
\end{equation}
where $D_\mu = \partial_\mu -iqA_\mu$ and the gauge as $A_t=-  Ex$. 
We solve this equation using separation of variables:
\begin{equation}
\Phi(t,x,\phi) = e^{-i\omega t} H_L(\theta) R(x),
\end{equation}
where $H_L(\theta)$ is the eigenfunction of the angular Laplacian on $S^1$, satisfying $\partial_\theta^2 H_L=-L^2 H_L$. We can write the full radial equation as:
\begin{equation}
\label{eq:radial_equation}
\begin{aligned}
x^{2} R''(x) + 2x R'(x)
&+ \Biggl[
    \frac{\ell^{4}}{4x^{2}} \bigl(\omega + q E x \bigr)^{2} \\
&\qquad
    - \frac{\ell^{2}}{2} \left( m^{2} + \frac{L^{2}}{\gamma^{2}} \right)
  \Biggr] R(x) = 0 .
\end{aligned}
\end{equation}
Which we write as 
\begin{equation} \label{radialeq.1}
    R^{\prime \prime}(x) + \frac{2}{x} R^{\prime}(x) + \left[ \frac{A}{x^4} + \frac{B}{x^3} + \frac{C}{x^2} \right] R(x)=0,
\end{equation}
where $A,B,$ and $C$ are given by:
\begin{align} \label{ABCcoef}
    A &= \frac{\ell^4 \omega^2}{4} \\
    B & =- \frac{\ell^4 \omega q E}{2} \\
    C &= \frac{\ell^4 q^2 E^2}{4} - \frac{\ell^2}{2} \left( m^2 + \frac{L^2}{\gamma^2} \right)
\end{align}
We use the following coordinate transformation
\begin{equation}
    z=\frac{k}{x},
\end{equation}
From which we find that:
\begin{align} \label{derivatives}
    \frac{dz}{dx} &= - \frac{z^2}{k} \\
    R^{\prime} (x)&= \frac{dR}{dz} \frac{dz}{dx} = -\frac{z^2}{k} R^\prime(z) \\
    R^{\prime \prime} (x) & =\frac{d}{dz} \left( -\frac{z^2}{k} R^\prime(z) \right) =\frac{z^4}{k^2} R^{\prime \prime} (z) + \frac{2z^3}{k^2} R^\prime(z).
\end{align}
Substituting eqs.\eqref{derivatives}  into the \eqref{radialeq.1}
we can write
\begin{equation}
    R^{\prime \prime} (z) + \left[ \frac{A}{k^2} + \frac{B}{k} \frac{1}{z} + \frac{C}{z^2} \right] R(z)=0,
\end{equation}
which we will math with Whittaker equation
\begin{equation} \label{radialeqWhittaker}
\partial_z^2 R(z) + \left[ -\frac{1}{4} + \frac{ \kappa}{z} + \frac{1/4 -  \mu^2}{z^2} \right] R(z) = 0.
\end{equation} 
We can write $\kappa$ and $\mu$ coefficients in temrs of $A, B$ and $C$ coefficients \eqref{ABCcoef} and we find:
\begin{align}
    k&=   i \ell^2 \omega  \\
    \kappa &=  -\frac{ i \ell^2 qE}{2} \\
    \mu^2 &=  \frac{1}{4}- C. 
\end{align}
From these parameters, we can find that Schwinger pair production is on when $\mu^2 < 0$ or $C > \frac{1}{4}$ from which we find the critical electric field when Schwinger pair production happens.
\begin{equation} \label{critialelectricfield}
    E_{\rm crit} = \frac{1}{\mid q \mid \ell^2} \sqrt{1+2\ell^2 \left( m^2 + \frac{L^2}{\gamma^2} \right)}
\end{equation}
A general solution of the radial equation \eqref{eq:radial_equation} is given by:
\begin{equation}
    R(x)=C_1 M_{ \kappa, \mu}\left(\frac{k}{x}\right) + C_2 W_{ \kappa, \mu}\left(\frac{k}{x}\right) 
\end{equation}
\subsection{IR asymptotics (horizon)}
Now we impose boundary conditions and for the IR we have. The large $\mid z \mid (x \to 0)$ horizon of $W$ and $M$ functions are given by:
\begin{equation}
\begin{split}
     & W_{\kappa,\mu}(z) \sim \exp\!\left(-\frac{z}{2}\right) z^{\kappa}
     \\
    & W_{\kappa,\mu}\!\left(\frac{k}{x}\right)
    \sim \exp\!\left(-\frac{k}{2x}\right)\left(\frac{k}{x}\right)^{\kappa}.
    \end{split}
\end{equation}
\begin{equation}
\begin{split}
    &M_{\kappa,\mu}(z) \sim{}
    \frac{\Gamma(1+2\mu)}{\Gamma\!\left(\tfrac{1}{2}+\mu-\kappa\right)}
    \exp\!\left(+\frac{z}{2}\right) z^{-\kappa} \\
    &+
    \frac{\Gamma(1+2\mu)}{\Gamma\!\left(\tfrac{1}{2}+\mu+\kappa\right)}
    \exp\!\left(-\frac{z}{2}\right) z^{\kappa}
\end{split}
\end{equation}
\begin{equation}
    \begin{split}
M_{\kappa,\mu}\!\left(\frac{k}{x}\right) & \sim
\frac{\Gamma(1+2\mu)}{\Gamma\!\left(\tfrac{1}{2}+\mu-\kappa\right)}
\exp\!\left(+\frac{k}{2x}\right)\left(\frac{k}{x}\right)^{-\kappa}
 \\
& +\frac{\Gamma(1+2\mu)}{\Gamma\!\left(\tfrac{1}{2}+\mu+\kappa\right)}
\exp\!\left(-\frac{k}{2x}\right)\left(\frac{k}{x}\right)^{\kappa}.
\end{split}
\end{equation}
Then for $k =-i \ell^2 \omega$, the two phases $\exp{\left( \mp \frac{k}{2x} \right)}$ are oscillatory and the infalling IR mode is the one proportional to $\exp{\left( - \frac{k}{2x} \right)}$. Therefore from it IR boundary condition we have
\begin{equation}
R_{\rm IR}(x) \propto W_{\kappa,\mu}\!\left(\frac{k}{x}\right)
\sim \exp\!\left(+\frac{i \ell^2 \omega}{2x}\right)\left(\frac{k}{x}\right)^{\kappa},
\quad x\to 0.
\end{equation}

\subsection{UV Asymptotics (\texorpdfstring{$\mathrm{AdS}_2$}{AdS₂} Boundary)}
We expand $W_{\kappa,\mu}$ for small $z$:
\begin{equation}
\begin{split}
    W_{\kappa,\mu}(z) \;=\;& 
    \frac{\Gamma(2\mu)}{\Gamma\!\left(\tfrac{1}{2}+\mu-\kappa\right)}\,
    z^{\frac{1}{2}-\mu}
    \\
    &+
    \frac{\Gamma(-2\mu)}{\Gamma\!\left(\tfrac{1}{2}-\mu-\kappa\right)}\,
    z^{\frac{1}{2}+\mu}
    + \mathcal{O}\!\left(z^{\frac{3}{2}\pm\mu}\right).
\end{split}
\end{equation}
and in terms of our $x$-coordinate we have
\begin{equation} \label{UVradial}
\begin{split}
R_{\rm UV}(x) = C_1 \bigg[ &
\underbrace{
\frac{\Gamma(2\mu)}{\Gamma\!\left(\tfrac{1}{2}+\mu-\kappa\right)}
k^{\frac{1}{2}-\mu}
}_{A_{-}(\kappa,\mu)}
\,x^{-\frac{1}{2}+\mu} \\
&+
\underbrace{
\frac{\Gamma(-2\mu)}{\Gamma\!\left(\tfrac{1}{2}-\mu-\kappa\right)}
k^{\frac{1}{2}+\mu}
}_{A_{+}(\kappa,\mu)}
\,x^{-\frac{1}{2}-\mu}
+\cdots \bigg],
\end{split}
\end{equation}
where we can write a more compact form as 
\begin{equation}
    R_{\rm UV}(x) \sim A_{-} x^{-\Delta_-} +  A_+ x^{-\Delta_+}.
\end{equation}
In the BF-violating regime the near-boundary exponent becomes imaginary,
\begin{equation}
\mu = i b, \qquad b\in\mathbb{R}.
\end{equation}
The two UV falloffs then take the form
\begin{equation}
x^{-\Delta_\pm} =
x^{-\frac{1}{2}\mp i b} = x^{-\frac{1}{2}} e^{\mp i b \log x},
\qquad \Delta_\pm = \frac{1}{2}\pm i b .
\end{equation}
Thus both branches have the same power-law magnitude, $x^{-1/2}$, but differ by an oscillatory logarithmic phase.
For the ingoing solution selected at the horizon, the coefficient of the slower falloff is
\begin{equation}
A_- = C_{\rm IR} \frac{\Gamma(2\mu)} {\Gamma\left(\frac{1}{2}+\mu-\kappa\right)} k^{\frac{1}{2}-\mu}.
\end{equation}
In standard quantization, a source-free UV boundary condition would require $A_-=0$. Since the Gamma function has no zeros, this can happen only when the denominator has a pole,
\begin{equation}
\frac{1}{2}+\mu-\kappa=-n, \qquad n\in\mathbb{Z}_{\ge0}.
\label{polecondition}
\end{equation}
However, in the BF-violating regime both $\mu$ and $\kappa$ are purely imaginary. Therefore the quantity
\begin{equation}
\frac{1}{2}+\mu-\kappa
\end{equation}
has real part $1/2$ and cannot lie on the negative real axis. Hence the source-free Dirichlet-type condition cannot be satisfied by the purely ingoing solution in this regime.\newline
Instead, in the BF-violating ($\mathrm{AdS}_2$ regime one must impose a mixed, self-adjoint UV boundary condition. Such a boundary condition fixes the relative phase, or equivalently the ratio, of the two logarithmically oscillatory branches. For the purely ingoing Whittaker solution this ratio is
\begin{equation}
\frac{A_+}{A_-} = k^{2\mu} \frac{\Gamma(-2\mu)} {\Gamma(2\mu)} \frac{
\Gamma\left(\frac{1}{2}+\mu-\kappa\right)}{\Gamma\left(\frac{1}{2}-\mu-\kappa\right)} .
\end{equation}
This is the self-adjoint, or mixed, boundary condition for the BF-violating $\mathrm{AdS}_2$ throat. The spectrum is then determined by equating this expression to the UV parameter that specifies the mixed boundary condition.\newline
When the electric field exceeds the critical value $E_{\rm crit}$, the effective mass of the scalar in the $\mathrm{AdS}_2$ region drops below the BF bound. The scaling dimensions become complex,
\begin{equation}
\Delta_\pm = \frac{1}{2}\pm i b ,
\label{scalingdim}
\end{equation}
and the UV wavefunctions become logarithmically oscillatory. The interpretation of the BF bound in $\mathrm{AdS}_2$ \cite{Breitenlohner:1982bm,Breitenlohner:1982jf} differs in an important way from higher-dimensional cases.\footnote{I thank Zohar Komargodski and Mark Mezei for insightful discussions on the interpretation of the BF bound and its relation to emergent infrared dynamics in $\mathrm{AdS}_2$.}
In $\mathrm{AdS}_{d+1}$, the BF bound marks the point at which the two independent falloffs degenerate, while in the alternative-quantization window the two possible quantizations correspond to distinct boundary conditions and operator dimensions. In the special case of $\mathrm{AdS}_2$, the saturation point gives an elementary scaling dimension $\Delta=1/2$, and the associated bilinear deformation is marginal. This feature underlies the emergence of nontrivial infrared dynamics and has been extensively discussed in the context of holographic quantum criticality and semi-local quantum liquids \cite{iqbal2010quantum,faulkner2011emergent,Iqbal:2011aj}.\newline
From this perspective, the instability observed here should be understood as a dynamical realization of this general mechanism. The violation of the $\mathrm{AdS}_2$ BF bound signals the onset of Schwinger pair production in the near-horizon throat and the appearance of a charged scalar mode. At the level of the fixed-background analysis, this mode is the linear precursor of a screening cloud. Related phenomena have also been discussed in the context of scalar clouds and long-lived excitations around black holes \cite{Hod:2012px,hod2015extremal,Herdeiro:2014goa,Degollado:2014vsa}.\newline
Physically, the electric field is strong enough to pull virtual charged pairs out of the vacuum. The negatively charged particle is accelerated toward the horizon, while the positively charged one is driven outward, producing an outgoing charged response in the UV. Geometrically, the throat behaves like a capacitor near its breakdown voltage: for $E>E_{\rm crit}$ charged pairs are continually produced in the throat. A finite screening cloud, however, requires including the Maxwell and gravitational backreaction generated at order $|\Phi|^2$.
\begin{figure}[htb]
\centering
\includegraphics[width=0.7\linewidth]{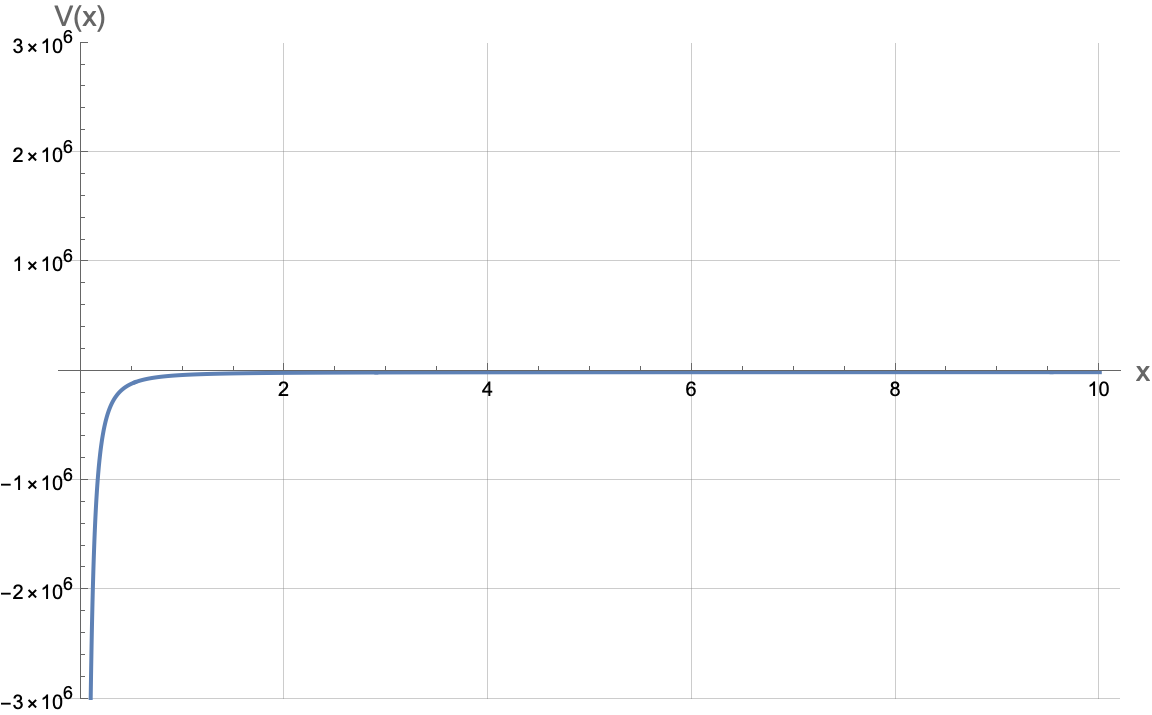}
\caption{Effective potential $\mathcal{V}_{\rm eff}(x)$ as a function of the radial coordinate $x$. The horizon is at $x\to0$, while the $\mathrm{AdS}_2$ boundary is at $x\to\infty$.}
\label{fig:eff_potential}
\end{figure}

It is useful to rewrite the radial equation in a Schrödinger-like form by removing the first-derivative term. Defining
\begin{equation}
R(x)=x^{-1/2}\psi(\log x),
\end{equation}
one finds
\begin{equation}
\frac{d^2\psi}{d(\log x)^2} - \mathcal{V}_{\rm eff}(x)\psi=0,
\end{equation}
where
\begin{equation}\label{eq:full_potential} \begin{split} \mathcal{V}(x)= & \underbrace{\Bigg[\, \frac{1}{4} -\frac{\ell^{4}(qE)^{2}}{4} +\frac{\ell^{2}}{2}\,m_{\mathrm{eff}}^{2}\,\Bigg]}_{\text{constant (BF part)}} \\ & +\underbrace{\frac{\ell^{4}\,\omega\,qE}{2}\,\frac{1}{x}}_{\text{Coulomb tilt}} -\underbrace{\frac{\ell^{4}\,\omega^{2}}{4}\,\frac{1}{x^{2}}}_{\text{near-horizon term}}\,. \end{split} \end{equation}
Thus the asymptotic sign of $\mathcal{V}_{\rm eff}$ determines whether the $\mathrm{AdS}_2$ BF bound is respected or violated:
\begin{equation}
\mu^{2}>0 \quad \Longleftrightarrow \quad C<\frac{1}{4} \quad \Longleftrightarrow \quad
E<E_{\rm crit},
\end{equation}
while
\begin{equation}
\mu^{2}<0 \quad \Longleftrightarrow \quad C>\frac{1}{4} \quad \Longleftrightarrow \quad E>E_{\rm crit}.
\end{equation}
The three terms in \eqref{eq:full_potential} have distinct physical roles:
\begin{itemize}
\item The constant term is the asymptotic BF contribution and controls the UV scaling dimensions in the $\mathrm{AdS}_2$ throat.
\item The $1/x$ term is a Coulomb tilt generated by the interference between the frequency $\omega$ and the electric coupling $qE$.
\item The $1/x^2$ term dominates near the horizon and is always attractive for real $\omega$.
\end{itemize}
Near the horizon, $x\to0$, the potential behaves as
\begin{equation}
\mathcal{V}_{\rm eff}(x)
\sim -\frac{\ell^{4}\omega^{2}}{4x^{2}},
\end{equation}
so $\mathcal{V}_{\rm eff}<0$ and the radial solutions are oscillatory. In the subcritical regime, $E<E_{\rm crit}$, the asymptotic value is positive,
\begin{equation}
\mathcal{V}_{\rm eff}(x\to\infty)=\mu^2>0,
\end{equation}
and the point where $\mathcal{V}_{\rm eff}(x)=0$ acts as a turning point between the oscillatory near-horizon region and the exponentially suppressed UV region.\
In the supercritical regime, $E>E_{\rm crit}$, the asymptotic value becomes negative,
\begin{equation}
\mathcal{V}_{\rm eff}(x\to\infty)=\mu^2<0.
\end{equation}
The UV behavior is then no longer exponentially suppressed; instead, the two branches become logarithmically oscillatory,
\begin{equation}
R(x)\sim x^{-1/2}e^{\mp ib\log x},
\qquad
\mu=ib.
\end{equation}
This is the potential-language manifestation of the $\mathrm{AdS}_2$ BF-bound violation and of the onset of the charged scalar instability.

\section{\label{sec:level4}Backreaction}
We now address the gravitational and Maxwell backreaction of the scalar
cloud. The fixed-background analysis of Sec.~\ref{sec:level3}
determines the linear onset of the instability, but a condensate of
finite amplitude sources both the gauge field and the metric. We
therefore consider the full Einstein--Maxwell--scalar system,
\begin{equation}
\begin{aligned}
S=&\frac{1}{16\pi G} \int d^3x\,\sqrt{-g} \left( R+\frac{2}{\ell^2} \right) \\
& +\int d^3x\,\sqrt{-g} \bigg[ -\frac14 F_{\mu\nu}F^{\mu\nu} -\frac12
\left(D_\mu\Phi\right)^* D^\mu\Phi -\frac12 m^2|\Phi|^2
\bigg].
\end{aligned}
\end{equation}
Variation with respect to the gauge field gives
\begin{equation}
\nabla_\mu F^{\mu\nu}=J^\nu,
\end{equation}
where
\begin{equation}
J^\nu = \frac{iq}{2}
\left[ \Phi^*D^\nu\Phi - \Phi\left(D^\nu\Phi\right)^*
\right].
\end{equation}
Variation with respect to the metric gives
\begin{equation}
G_{\mu\nu} -\frac{1}{\ell^2}g_{\mu\nu} =
8\pi G \left( T_{\mu\nu}^{(A)} + T_{\mu\nu}^{(\Phi)}
\right),
\end{equation}
with
\begin{align}
T_{\mu\nu}^{(A)}
&= F_{\mu\rho}F_\nu{}^\rho -\frac14 g_{\mu\nu}F_{\rho\sigma}F^{\rho\sigma}, \\
T_{\mu\nu}^{(\Phi)}
 & = \frac12
\left[ \left(D_\mu\Phi\right)^* \left(D_\nu\Phi\right) + \left(D_\nu\Phi\right)^* \left(D_\mu\Phi\right) \right]
\nonumber\\
& \quad -\frac12 g_{\mu\nu}
\left[ \left(D_\rho\Phi\right)^*D^\rho\Phi + m^2|\Phi|^2 \right].
\end{align}
We use the static, circularly symmetric ansatz
\begin{equation}
ds^2 = -e^{-\chi(\rho)}f(\rho)\, dt^2 +
\frac{d\rho^2}{f(\rho)} + \rho^2d\theta^2,
\end{equation}
together with
\begin{equation}
A=A_t(\rho)\,dt,
\qquad
\Phi=\psi(\rho).
\end{equation}
The scalar is taken to be static and axisymmetric. Its phase can be
removed by a gauge transformation, so that $\psi(\rho)$ may be chosen
real. The Klein-Gordon equation then becomes
\begin{equation}
\psi'' + \left( \frac{f'}{f} + \frac{1}{\rho} - \frac{\chi'}{2} \right)\psi' + \left( \frac{q^2e^\chi A_t^2}{f^2} - \frac{m^2}{f}
\right)\psi = 0.
\end{equation}
The Maxwell equation reduces to
\begin{equation}
A_t'' + \left( \frac{1}{\rho} + \frac{\chi'}{2} \right)A_t' -
\frac{q^2\psi^2}{f}A_t = 0.
\end{equation}
The difference between the $tt$ and $\rho\rho$ components of the
Einstein equations gives
\begin{equation}
\chi' = -16\pi G\,\rho
\left[ (\psi')^2 + \frac{q^2e^\chi A_t^2}{f^2}\psi^2
\right],
\end{equation}
while the $tt$ equation gives
\begin{equation}
f' - \frac{2\rho}{\ell^2} + 8\pi G\,\rho
\left[ e^\chi(A_t')^2 + f(\psi')^2 + \frac{q^2e^\chi A_t^2}{f}\psi^2
+ m^2\psi^2 \right] = 0.
\end{equation}
Setting $\psi=0$ and fixing the asymptotic time normalization so that
$\chi=0$, we recover the charged-BTZ solution
\eqref{fullmetric}.
We next introduce a small scalar amplitude. Since the scalar current
and stress tensor are quadratic in the scalar profile, neither the
Maxwell field nor the metric receives a correction at order
$\epsilon$. The first corrections therefore appear at order
$\epsilon^2$. We expand
\begin{align}
\psi &= \epsilon\psi_1 + \epsilon^3\psi_3 +\cdots, \\
A_t &= A_0 + \epsilon^2A_2 +\cdots, \\
f & = f_0 + \epsilon^2f_2 +\cdots, \\
\chi &= \epsilon^2\chi_2 +\cdots.
\end{align}
At zeroth order, the Maxwell and Einstein equations reduce to
\begin{align}
A_0'' + \frac{1}{\rho}A_0' &= 0, \\
f_0' - \frac{2\rho}{\ell^2} + 8\pi G\,\rho(A_0')^2 &= 0.
\end{align}
These equations reproduce the charged-BTZ background. In particular,
$A_0'=Q/\rho$, while the second equation gives
\begin{equation}
f_0(\rho) = \frac{\rho^2}{\ell^2} + C_0 - 8\pi GQ^2
\log\left(\frac{\rho}{\ell}\right),
\end{equation}
which is equivalent to Eq.~\eqref{fullmetric} after expressing the
integration constant $C_0$ in terms of the mass.\newline
At order $\epsilon$, the scalar obeys the linear charged
Klein--Gordon equation on the fixed charged-BTZ background,
\begin{equation}
\psi_1'' +
\left( \frac{f_0'}{f_0} + \frac{1}{\rho} \right)\psi_1'
+ \left( \frac{q^2A_0^2}{f_0^2} - \frac{m^2}{f_0} \right)\psi_1 =0.
\end{equation}
At order $\epsilon^2$, the scalar sources both the Maxwell field and
the geometry:
\begin{equation}
\chi_2'
=
-16\pi G\,\rho
\left[
(\psi_1')^2
+
\frac{q^2A_0^2}{f_0^2}\psi_1^2
\right].
\end{equation}
\begin{equation}
A_2'' + \frac{1}{\rho}A_2' + \frac{1}{2}\chi_2'A_0' - \frac{q^2\psi_1^2}{f_0}A_0 = 0.
\end{equation}
\begin{equation}
\begin{split}
f_2' = -8\pi G\,\rho
\Bigg[ &2A_0'A_2' + \chi_2(A_0')^2 \\ &+ f_0(\psi_1')^2 +
\frac{q^2A_0^2}{f_0}\psi_1^2 \\ &+ m^2\psi_1^2 \Bigg].
\end{split}
\end{equation}

\begin{figure}[htbp]
\centering
\includegraphics[width=0.95\linewidth]{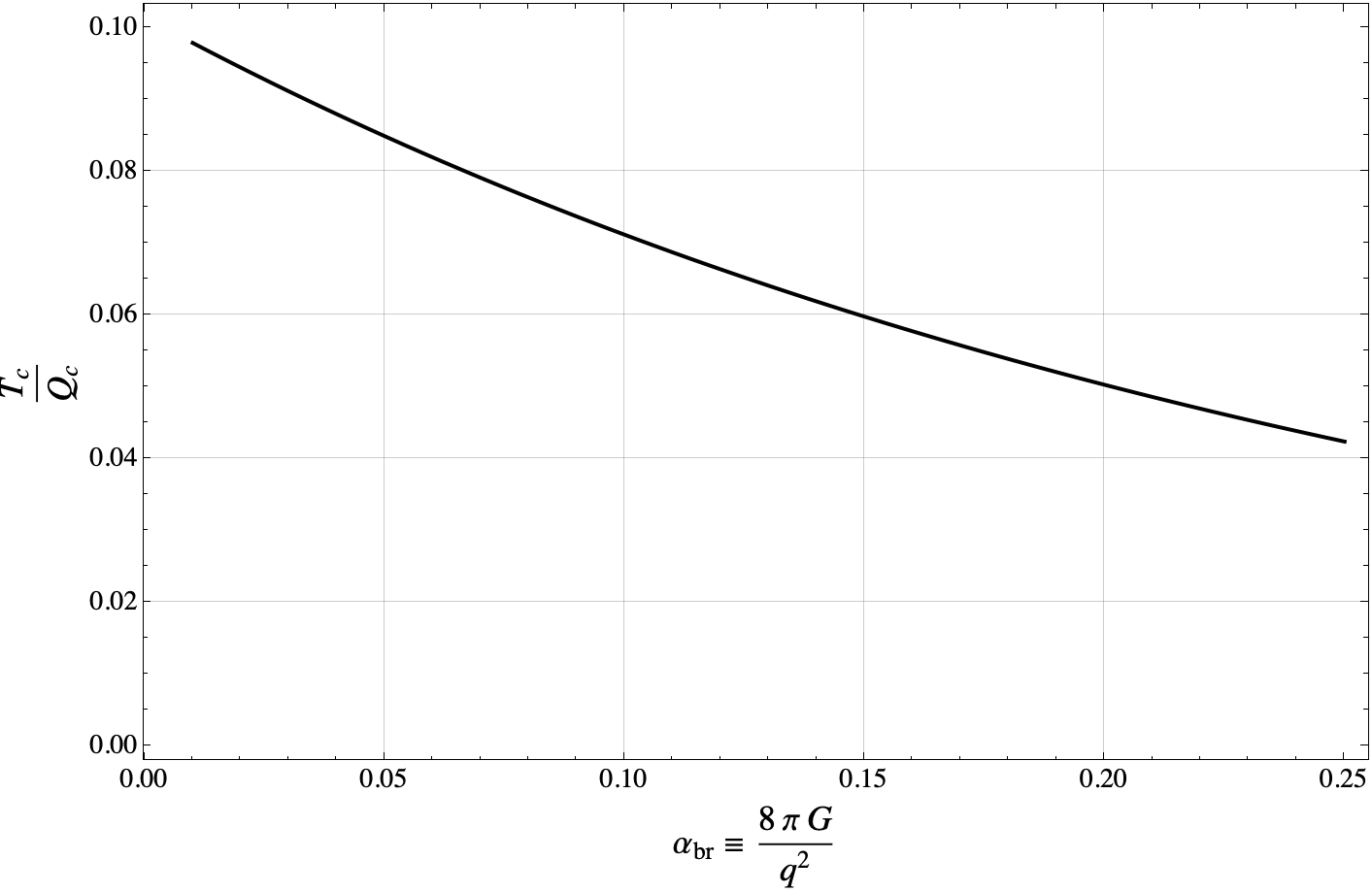}
\caption{Critical dimensionless temperature $T_c/Q_c$ as a function of the effective backreaction parameter $\alpha_{\rm br}=(8\pi G)/q^2$, with $q=2$. The monotonic decrease shows that stronger backreaction suppresses the onset of the charged-scalar instability.}
\label{fig:backreactionTc}
\end{figure}

These equations show explicitly that the scalar cloud necessarily sources both the Maxwell field and the geometry at order $\epsilon^2$. Therefore, the fixed charged-BTZ background used in the linear analysis is valid only for determining the onset of the instability. A finite-amplitude scalar configuration must instead be described by the full coupled Einstein--Maxwell--scalar equations.
We solve the full system numerically by shooting from the regular horizon to the asymptotic $\mathrm{AdS}_3$ boundary, following the standard strategy used in holographic superconductor constructions \cite{Hartnoll:2008vx,Hartnoll:2008kx} and in backreacted BTZ superconductors \cite{Liu:2011fy}. The detailed near-horizon and UV expansions are given in Appendix~A. Here we summarize only the data required for the shooting problem.
At large $\rho$, the scalar behaves as
\begin{equation}
\psi(\rho) = \psi_- \rho^{-\Delta_-} +
\psi_+ \rho^{-\Delta_+} +\cdots ,
\end{equation}
where
\begin{equation}
\Delta_{\pm} = 1 \pm \sqrt{1+m^2\ell^2}.
\end{equation}
We work in standard quantization and impose the source-free boundary condition
\begin{equation}
\psi_-=0.
\end{equation}
With this choice, $\psi_+$ is proportional to the expectation value of the dual operator and will be used as the condensate \cite{Hartnoll:2008vx,Hartnoll:2008kx,Liu:2011fy}.
The gauge field and the metric behave asymptotically as
\begin{equation}
A_t(\rho) =
Q \log{\frac{\rho}{\ell}} +
\mu + \cdots ,
\end{equation}
and
\begin{equation}
e^{-\chi(\rho)} f(\rho) = \frac{\rho^2}{\ell^2} -
8 \pi G Q^2 \log{\frac{\rho}{\ell}} -8GM + \cdots .
\end{equation}
\begin{figure*}
    \centering
    \includegraphics[width=0.7\linewidth]{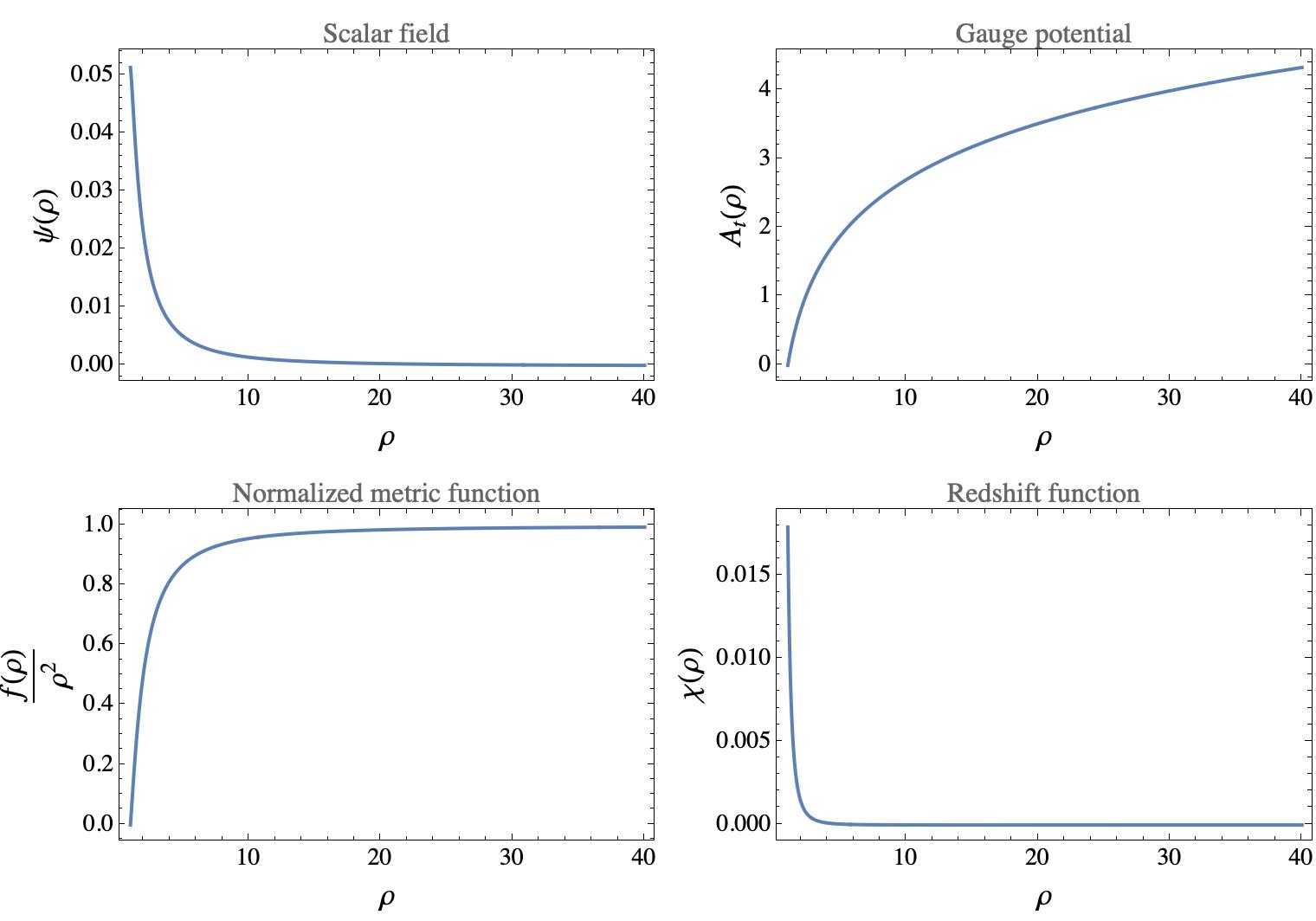}
    \caption{Representative radial profiles of the fully backreacted, source-free, node-free solution. The scalar field is regular at the horizon and decays toward the asymptotic AdS$_3$ boundary. The gauge potential satisfies $A_t(\rho_h)=0$, while the normalized metric function obeys $f(\rho)/\rho^2\to 1$ and the redshift function is normalized such that $\chi(\infty)=0$.}
    \label{radialprofiles}
\end{figure*}
The profiles confirm that the nonlinear branch defines a regular asymptotically AdS$_3$ black-hole solution. In particular, the scalar configuration is node-free and satisfies the source-free boundary condition, while the gauge field and metric functions obey the required horizon and ultraviolet boundary conditions. Thus, once the condensate acquires finite amplitude, the Maxwell field and geometry adjust self-consistently rather than remaining fixed to the charged-BTZ background. As a numerical consistency check, the maximum absolute residuals of the scalar, Maxwell, and two independent Einstein equations over the integration domain are below $10^{-5}$ for the representative solution. The coefficient $Q$ of the logarithmic Maxwell mode sets the natural electric scale, while $M$ is extracted from the constant term in the asymptotic metric after subtracting the logarithmic electric contribution.
At the horizon, we introduce
\begin{equation}
y=\rho-\rho_h
\end{equation}
and impose regularity,
\begin{equation}
f(\rho_h)=0,
\qquad
A_t(\rho_h)=0.
\end{equation}
The fields admit Taylor expansions of the form
\begin{align}
f(\rho)&=f_1 y+\cdots \\
A_t(\rho)&=E_h y+\cdots \\
\psi(\rho)&=\psi_h+\cdots \\
\chi(\rho)&=\chi_h+\cdots .
\end{align}
All higher coefficients are fixed recursively by the equations of motion. In practice, $\psi_h$ and $E_h$ are used as shooting parameters. The residual freedom to rescale time is fixed by imposing
\begin{equation}
\chi(\infty)=0.
\end{equation}
With this normalization, the Hawking temperature is
\begin{equation}
T = \frac{e^{-\chi(\rho_h)/2} f_1}{4\pi}.
\end{equation}
Since the operator dual to $\psi_+$ has scaling dimension $\Delta_+$, the coefficient $\psi_+$ has dimension $\Delta_+$. The charge parameter $Q$, appearing as the coefficient of the logarithmic Maxwell falloff, has the dimension of energy and therefore provides a natural normalization scale. We therefore use the dimensionless condensate
\begin{equation}
\frac{\psi_+}{Q^{\Delta_+}},
\end{equation}
and the dimensionless temperature
\begin{equation}
\frac{T}{Q}.
\end{equation}
The critical point is characterized by the ratio
\begin{equation}
\frac{T_c}{Q_c},
\end{equation}
and we study the condensate as a function of the reduced distance from the critical point,
\begin{equation}
1-\frac{T/Q}{T_c/Q_c}.
\end{equation}
For the representative numerical choice
\begin{equation}
\ell=1, \qquad 8\pi G=1, \qquad q=2, \qquad m^2=-\frac14,
\end{equation}
we first determine the onset of the instability by solving the linear scalar equation on the normal charged-BTZ branch. With $\rho_h=1$, the zero-node source-free mode appears at
\begin{equation}
E_h^{(c)} \simeq 1.1722683717 .
\end{equation}
On the normal branch one has
\begin{equation} 
Q_c=E_h^{(c)},
\end{equation}
and
\begin{equation}
T_c = \frac{2-\left(E_h^{(c)}\right)^2}{4\pi}.
\end{equation}
Thus the critical dimensionless temperature is
\begin{equation}
\frac{T_c}{Q_c} \simeq 0.04248 .
\end{equation}
For $T/Q<T_c/Q_c$, we find a family of fully backreacted solutions satisfying the source-free boundary condition $\psi_-=0$, with a nonzero condensate $\psi_+\neq0$. All solutions displayed below are zero-node configurations and therefore belong to the fundamental scalar branch rather than to radially excited branches. The condensate grows monotonically as $T/Q$ is lowered below $T_c/Q_c$. Near the onset, the numerical data are well described by
\begin{equation}
\frac{\psi_+}{Q^{\Delta_+}} = A
\left[ 1-\frac{T/Q}{T_c/Q_c} \right]^{\beta}.
\label{fitform}
\end{equation}
The instability is triggered by the violation of the effective
$\rm AdS_2$ Breitenlohner-Freedman bound in the extremal charged-BTZ
near-horizon region. The nonlinear solutions constructed here show
that this instability extends to a finite-temperature hairy branch
below the critical ratio $T_c/Q_c$. The numerical branch is restricted
to the near-critical regime: for the parameter ranges studied, the
solutions extend approximately $3\%-4\%$ below the corresponding
critical temperature. We therefore do not claim to determine the exact
zero-temperature endpoint of the instability. For the representative choice above, a fit to the near-critical data gives
\begin{equation}
A \approx 1.303,\qquad \beta \approx 0.498,
\end{equation}
which is consistent with the mean-field value $\beta=1/2$. Fixing
$\beta=1/2$, the fit becomes
\begin{equation}
\label{fitfixed}
\frac{\psi_+}{Q^{\Delta_+}} =
1.308 \left[ 1-\frac{T/Q}{T_c/Q_c} \right]^{1/2}.
\end{equation}
This shows that the scalar cloud identified in the fixed-background
analysis continues into a genuine backreacted hairy charged-BTZ
solution. The geometry is not held fixed: both the Maxwell field and
the metric adjust consistently to the scalar condensate.
\begin{figure}[htbp]
    \centering
    \includegraphics[width=0.95\linewidth]{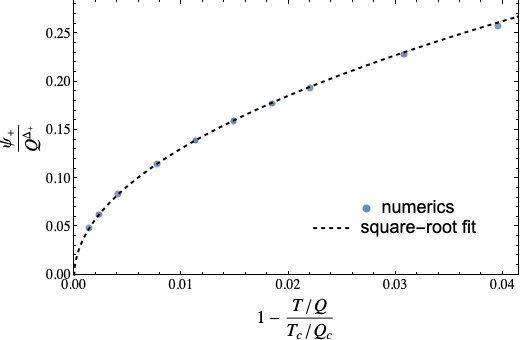}
    \caption{Condensate branch of the fully backreacted Einstein--Maxwell--scalar system for $\ell=1$, $8\pi G=1$, $q=2$, and $m^2=-1/4$. The scalar source coefficient $\psi_-$ is tuned to zero and all displayed solutions are zero-node configurations. The dashed curve is the near-critical square-root fit.}
    \label{fig:numerical}
\end{figure}
As a further consistency check, we vary the effective backreaction parameter
\begin{equation}
\alpha_{\rm br} =
\frac{8\pi G}{q^2}.
\end{equation}
The finite-temperature hairy solutions constructed above do not possess
an exact AdS$_2\times S^1$ near-horizon region, since their horizons are nonextremal.
The finite-temperature hairy solutions constructed above do not possess an exact AdS$_2\times S^1$ near-horizon region, since their horizons are nonextremal and $f(\rho)$ has a simple zero rather than the double zero required at extremality. To assess the zero-temperature limit, we considered an extremal near-horizon expansion with $f(\rho)\sim f_2(\rho-\rho_h)^2$, $A_t(\rho)\sim E_h(\rho-\rho_h)$, and $\psi(\rho)\sim \psi_0(\rho-\rho_h)^\lambda$. Regularity of the Maxwell and Einstein equations requires $\lambda>1/2$, whereas in the regime where the effective AdS$_2$ Breitenlohner-Freedman bound is violated the scalar exponents have real part $-1/2$. This indicates that the condensed zero-temperature endpoint cannot preserve the original regular AdS$_2\times S^1$ throat within this power-law ansatz. The infrared geometry or electric-field profile must therefore reorganize, and determining that endpoint lies beyond the near-critical finite-temperature analysis presented here.
For each value of $\alpha_{\rm br}$, we solve the linear scalar equation on the corresponding normal charged-BTZ background and determine the zero-node source-free mode. The resulting critical ratio $T_c/Q_c$ is shown in Fig.~\ref{fig:backreactionTc}. We find that $T_c/Q_c$ decreases monotonically as $\alpha_{\rm br}$ is increased. Thus, stronger gravitational backreaction suppresses the charged-scalar instability and makes condensation harder. This is the expected qualitative behavior in backreacted holographic superconductor models \cite{Hartnoll:2008kx,Liu:2011fy}.
We next construct the full nonlinear condensate branches for several values of $\alpha_{\rm br}$. In all cases, the scalar source is tuned to zero and the corresponding scalar profiles are node-free. The near-critical data are fitted by the general form in Eq.~\eqref{fitform}, and the fitted values are listed in Table~\ref{tab:backreaction-fit}.
\begin{table}[t]
\begin{ruledtabular}
\begin{tabular}{ccc}
$\alpha_{\rm br}=(8\pi G)/q^2$ & $A$ & $\beta$ \\
\hline
$0.050$ & $1.9769$ & $0.49799$ \\
$0.075$ & $1.8880$ & $0.49798$ \\
$0.125$ & $1.7119$ & $0.49795$ \\
$0.180$ & $1.5251$ & $0.49794$ \\
$0.250$ & $1.3047$ & $0.49793$ \\
\end{tabular}
\end{ruledtabular}
\caption{
Near-critical fit parameters for the nonlinear condensate branches.
The condensate is fitted as
$\psi_+/Q^{\Delta_+}
=A\left[1-(T/Q)/(T_c/Q_c)\right]^\beta$.
The exponent remains close to the mean-field value $\beta=1/2$,
while the amplitude decreases with increasing backreaction.
}
\label{tab:backreaction-fit}
\end{table}
The fitted exponent remains very close to the mean-field value $\beta=1/2$, while the amplitude decreases as the effective backreaction is increased. This confirms the same physical trend seen already in the linear onset calculation: stronger backreaction suppresses the condensate.
\begin{figure}[htbp]
\centering
\includegraphics[width=0.95\linewidth]{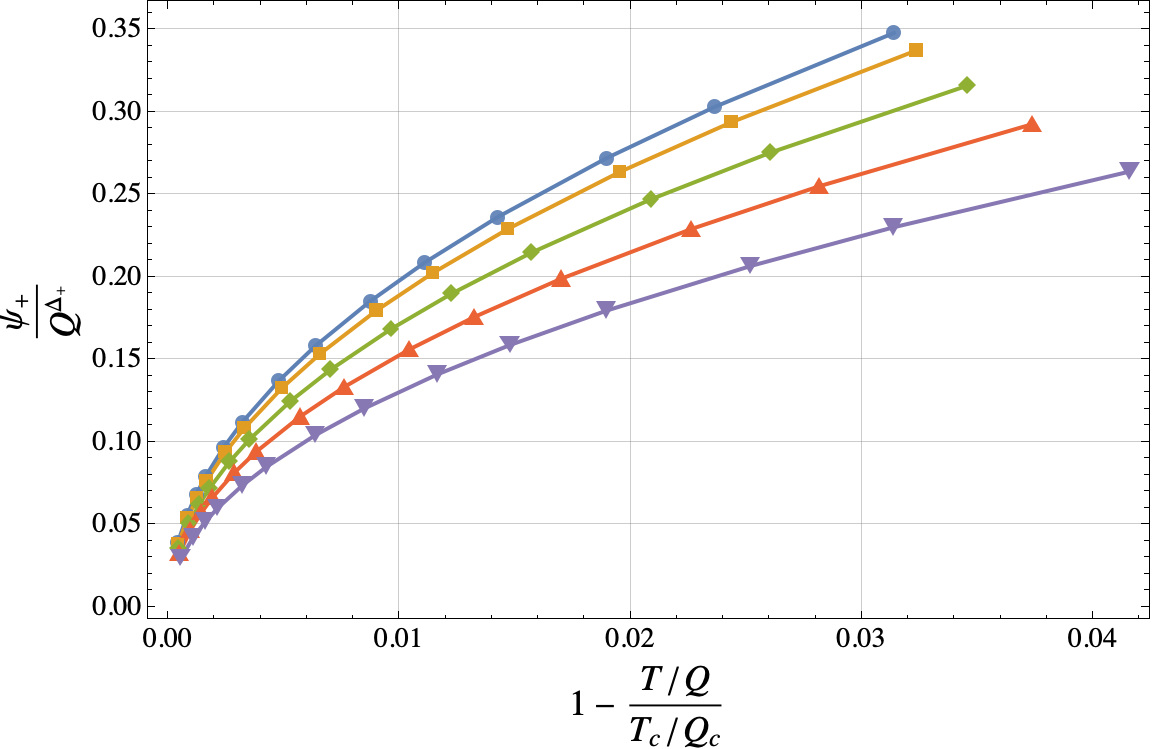}
\caption{Nonlinear condensate branches for several values of the effective backreaction parameter $\alpha_{\rm br}=(8\pi G)/q^2$, with $q=2$. The horizontal axis is the reduced distance from the critical point, $1-(T/Q)/(T_c/Q_c)$. Increasing $\alpha_{\rm br}$ suppresses the condensate amplitude, while the near-critical exponent remains consistent with $\beta=1/2$.}
\label{fig:nonlinearbranches}
\end{figure}

\section{\label{sec:level5}Including Boundary Terms}
The fully backreacted solutions constructed in Sec.~\ref{sec:level4} were obtained with standard asymptotic $\mathrm{AdS}_3$ boundary conditions. In this section we instead isolate the near-horizon $\mathrm{AdS}_2$ throat and discuss the boundary terms required to formulate a well-posed variational problem for the effective $\mathrm{AdS}_2$ scalar. This analysis should be viewed as the effective near-horizon interpretation of the scalar cloud, rather than as a replacement of the asymptotic $\mathrm{AdS}_3$ boundary conditions used in the nonlinear construction. It provides an interpretation of the scalar cloud in terms of mixed boundary conditions, double-trace flow, and defect screening.
As emphasized in Refs.~\cite{Ferreira:2017cta,Dappiaggi:2017pbe}, timelike AdS boundaries allow one to impose more general boundary conditions than purely Dirichlet ones. In particular, mixed or Robin boundary conditions correspond to allowing boundary degrees of freedom, sources, or absorbers to interact with the bulk field. In holography, the same structure appears when multi-trace operators are added to the dual theory. The associated mixed boundary conditions and double-trace renormalization-group flows have a standard interpretation in AdS/CFT and play an important role in defining consistent variational principles and holographic RG flows \cite{Witten:2001ua,Berkooz:2002ug,Hartman:2006dy,das2013double,Aharony:2015afa}.
We first switch from the coordinate $x$ used in Sec.~\ref{sec:level3} to a radial coordinate $r$ that vanishes at the $\mathrm{AdS}_2$ boundary,
\begin{equation}
r \equiv \frac{1}{x}.
\label{r-coord}
\end{equation}
Thus $r\to 0$ corresponds to the $mathrm{AdS}_2$ boundary, while $r\to\infty$ corresponds to the horizon. The UV expansion \eqref{UVradial} becomes
\begin{equation}
R(r) = A_- r^{\Delta_-} + A_+ r^{\Delta_+} +\cdots ,
\label{R-falloffs}
\end{equation}
where
\begin{equation}
\Delta_\pm=\frac12\pm\mu .
\end{equation}
For real $\mu>0$, $A_-$ multiplies the less rapidly decaying mode $r^{\Delta_-}$, while $A_+$ multiplies the more rapidly decaying mode $r^{\Delta_+}$. We introduce a small radial cutoff $r=r_0$, which plays the role of a UV wall in the effective $\mathrm{AdS}_2$ problem.\newline
After reducing over the rigid circle $S^1$ and focusing on a fixed angular harmonic $L$, the scalar dynamics in the throat is described by an effective charged scalar in $\mathrm{AdS}_2$, with mass $(m_{\rm eff}$. Up to an overall normalization, the bulk action is
\begin{equation}
S_{\rm bulk} =
\int d^2x\sqrt{-g_{(2)}}
\left[ -\frac12 (D_\mu\Phi)^*D^\mu\Phi
-\frac12 m_{\rm eff}^2|\Phi|^2 \right],
\label{AdS2-bulk-action}
\end{equation}
where $D_\mu=\nabla_\mu-iqA_\mu$. Varying the action and integrating by parts gives
\begin{equation}
\delta S_{\rm bulk}=
\text{(EOM terms)} +
\delta S_{\rm bdy}^{\rm bulk},
\end{equation}
with boundary contribution
\begin{equation}
\delta S_{\rm bdy}^{\rm bulk} =
\int_{r=r_0} dt\sqrt{-\gamma}\, n^r
\left[ (D_r\Phi)^*\delta\Phi + \text{c.c.}
\right].
\label{bulk-boundary-variation}
\end{equation}
Using the expansion \eqref{R-falloffs}, the leading near-boundary variation contains a divergent term of the form
\begin{equation}
\delta S_{\rm bdy}^{\rm bulk}
\supset C \, \Delta_- \, r_0^{-2\mu}
\left( A_-^*\delta A_- + \text{c.c.} \right) +\cdots ,
\end{equation}
where $C$ is a positive normalization constant determined by the induced metric, and the ellipsis denotes finite and subleading terms involving $A_\pm$ and their variations. This divergent contribution is local in the boundary data and is removed by adding the standard boundary counterterm
\begin{equation}
S_{\rm bdy}^{(1)} = -\frac{1-2\mu}{2}
\int_{r=r_0}dt\sqrt{-\gamma}\,|\Phi|^2 .
\label{first-boundary-term}
\end{equation}
The combined variation then has the finite leading form
\begin{equation}
\delta\left(S_{\rm bulk}+S_{\rm bdy}^{(1)}\right)
= 2\mu C
\int d\omega
\left( A_+^*\delta A_- + \text{c.c.} \right)
+\cdots .
\label{renormalized-standard-variation}
\end{equation}
Therefore a standard, Dirichlet-type variational principle is obtained by fixing
\begin{equation}
\delta A_-(t)=0 .
\end{equation}
In holographic language $A_-$ is the source, while $A_+$ is determined dynamically and gives the response. Setting $A_-=0$ is the source-free condition; it is a choice of state, not the definition of the variational principle. In this standard quantization, the dual defect operator has scaling dimension
\begin{equation}
\Delta_+=\frac12+\mu .
\end{equation}
If one insists on strictly local gauge-invariant defect operators in the effective throat theory, the charged scalar itself is not the simplest local observable; instead one should consider neutral composites. The simplest such operator is the bilinear \cite{Aharony:2023amq,Aharony:2022ntz}
\begin{equation}
\mathcal{O}_{\rm std}
\sim
\Phi^*\Phi,
\qquad
\widehat{\Delta}_+
= 2\Delta_+ = 1+2\mu .
\end{equation}
Within the window $0<\mu<1/2$, this bilinear is irrelevant in the standard quantization. Thus adding it as a local boundary interaction does not generate a nontrivial infrared instability in the subcritical $\mathrm{AdS}_2$ problem. In the decoupling limit $q\to0$, with $m_{\rm eff}=0$, the electric coupling is removed and one has $\mu\to 1/2$. The bilinear then has
\begin{equation}
\widehat{\Delta}_+=2,
\end{equation}
as expected for the ordinary $\mathrm{AdS}_2/\mathrm{CFT}_1$ vacuum without a charged defect.
The standard quantization described above treats $A_-$ as the source and $A_+$ as the response. In the window $0<\mu<1/2$, one may instead pass to the alternative quantization by performing the usual Legendre transform of the renormalized boundary action. Equivalently, one adds a finite boundary term chosen so that the variational principle fixes $A_+$ rather than $A_-$. In this quantization the finite boundary variation takes the schematic form
\begin{equation}
\delta S_{\rm alt} = -2\mu C
\int d\omega
\left(  A_-^*\,\delta A_+ + \text{c.c.} \right),
\label{alt-variation}
\end{equation}
where $C$ is a positive normalization constant depending on the normalization of the $\mathrm{AdS}_2$ action. Thus a well-posed variational principle is obtained by imposing
\begin{equation}
\delta A_+(t)=0 \qquad \text{at } r=r_0 .
\end{equation}
In the alternative quantization, $A_+$ plays the role of the source, while $A_-$ is dynamical and determines the expectation value. Setting $A_+=0$ corresponds to turning off the source in this quantization.
In the alternative quantization, the operator dual to the scalar has dimension
\begin{equation}
\Delta_-=\frac{1}{2}-\mu .
\end{equation}
The leading near-boundary coefficient $A_-$ is therefore interpreted as the expectation value of this operator. If we again restrict to neutral local composites in the effective defect description, the simplest gauge-invariant bilinear has scaling dimension
\begin{equation}
\widehat{\Delta}_{\rm alt} = 2\Delta_- = 1-2\mu .
\end{equation}
For $0<\mu<\frac{1}{2}$, this operator is relevant. Therefore perturbing the alternative defect CFT by this bilinear produces a double-trace deformation.\newline
In the bulk description this deformation is implemented by adding a second boundary term at the cutoff surface,
\begin{equation}
S_{\rm bdy}^{(2)} = -f_0 \int_{r=r_0}
dt\sqrt{-\gamma}\, r_0^{2\mu}
|\Phi|^2 .
\label{bdyterm2}
\end{equation}
The factor $r_0^{2\mu}$ is included so that the induced $|A_-|^2$ term remains finite as $r_0\to0$. At leading order, this term modifies the alternative-quantization variation to
\begin{equation}
\delta \left(S_{\rm alt}+S_{\rm bdy}^{(2)}\right)
= -2\mu C \int d\omega
\left[ A_-^* \delta\left(A_+-f A_-\right) + \text{c.c.} \right],
\label{mixed-variation}
\end{equation}
where $f$ is the dimensionless double-trace coupling at the cutoff scale, up to a convention-dependent normalization relating $f$ and $f_0$. The source in the deformed theory is therefore
\begin{equation}
J_f=A_+-fA_- .
\end{equation}
The source-free condition in the deformed theory gives the mixed, or Robin, boundary condition
\begin{equation}
A_+=f(r_0)A_- .
\label{mixed-bc}
\end{equation}
The ratio $A_+/A_-$ is a physical quantity and must be independent of the arbitrary cutoff $r_0$. Defining the field-theory RG scale by
\begin{equation}
\Lambda\sim \frac1{r_0},
\end{equation}
the running of the double-trace coupling is
\begin{equation}
\beta_f \equiv
\frac{df}{d\log\Lambda} = -2\mu f+f^2 .
\label{double-trace-beta}
\end{equation}
For $0<\mu<\frac12$, this flow has two fixed points,
\begin{equation}
f_*=0, \qquad f_*=2\mu .
\end{equation}
In this convention, these correspond respectively to the alternative and standard quantizations. This double-trace RG interpretation applies when $\mu$ is real. In the BF-violating regime, 
\begin{equation} 
\mu=ib, 
\end{equation}
the scaling dimensions become complex and the ordinary real fixed-point structure is lost. The appropriate boundary condition is then better described by a self-adjoint phase, or equivalently by the ratio $A_+/A_-$, rather than by a flow between two real RG fixed points.\newline
It is useful to interpret this boundary structure from the perspective of worldline effective field theory. In this language, a heavy charged object is represented by a Wilson line coupled to the bulk gauge field, as in effective descriptions of charged compact objects \cite{Goldberger:2004jt,Goldberger:2005cd}. The extremal charged BTZ black hole then acts as a static electrically charged defect, whose interaction with the near-horizon fields is encoded in boundary data at the ${\rm AdS}_2$ edge. Mixed boundary conditions and the associated double-trace flow describe the dynamical dressing of this defect by charged matter. In particular, the scalar cloud generated by Schwinger pair production can be interpreted as a screening cloud that renormalizes the effective charge of the defect. This is closely analogous to the renormalization of Wilson lines in scalar QED and related systems \cite{Aharony:2015afa,Aharony:2022ntz,Aharony:2023amq}, but here it is realized dynamically in the curved near-horizon geometry of an extremal charged BTZ black hole.\newline
Before turning to the BF-violating regime, let us briefly recall what happens below the electric threshold. When
\begin{equation}
\mu=\sqrt{\frac14-C}>0,
\qquad
C<\frac14,
\end{equation}
the two ${\rm AdS}_2$ falloffs have real dimensions
\begin{equation}
\Delta_\pm=\frac12\pm\mu .
\end{equation}
The near-horizon throat is then above the $(\mathrm{AdS}_2$ BF bound. The standard and alternative quantizations are both well defined in the window $0<\mu<1/2$, and the mixed boundary condition may be interpreted in terms of the real double-trace flow discussed above. Within the fixed-background effective analysis, the subcritical regime contains no intrinsic source-free charged-scalar instability. The normal extremal charged-BTZ throat remains the relevant scalar-free background, 
\begin{equation} 
\Phi=0, \qquad A_t(x)=-Ex, 
\end{equation} 
with $F_{tx}=E$. Thus, for $E<E_{\rm crit}$, the effective mass does not violate the $\mathrm{AdS}_2$ BF bound, Schwinger pair production is suppressed, and no scalar screening cloud is generated dynamically. The cloud studied below is therefore a genuinely supercritical phenomenon, tied to the violation of the electric BF threshold.

\subsection{BF Violation and the Phase Boundary Condition}
We now turn to the regime in which the near-horizon electric field exceeds the electric BF threshold. Equivalently,
\begin{equation}
\mu^2<0 \qquad\Longleftrightarrow\qquad C>\frac14,
\end{equation}
and we write
\begin{equation}
\mu=ib, \qquad b>0 .
\label{supercriticalregime}
\end{equation}
The two $\mathrm{AdS}_2$ dimensions then become complex,
\begin{equation}
\Delta_\pm = \frac12\pm ib .
\end{equation}
This complex scaling does not by itself signal an inconsistency of the full theory. Rather, it signals that the scalar-free defect saddle is unstable: the electric field has become strong enough that the effective scalar mass violates the $\mathrm{AdS}_2$ BF bound.\newline
In this regime the ordinary real double-trace flow between two fixed points is no longer the appropriate description. Instead, the two independent near-boundary modes are logarithmically oscillatory,
\begin{equation}
R(r) = A_- r^{1/2-ib} +
A_+ r^{1/2+ib} +\cdots .
\end{equation}
For a real radial profile, the coefficients may be chosen so that $A_+=A_-^*$. The solution can then be written as
\begin{equation}
R(r) = {\cal C}\sqrt{r}\,
\cos\left[ b\log\frac{r}{r_0} + \delta \right],
\label{scalarprofile}
\end{equation}
where the phase $\delta$ is fixed by the ratio $A_+/A_-$, up to the convention used to define the reference scale $r_0$. Thus, in the BF-violating regime, the UV boundary condition is better described by a phase, or self-adjoint extension, rather than by a real RG fixed point.\newline
This phase has a simple physical interpretation. The electric field produces charged pairs in the throat. The horizon absorbs infalling charge, while the UV boundary condition determines how the outgoing component is reflected back into the throat. Since the natural radial coordinate is logarithmic, the scalar mode accumulates a phase proportional to $b\log r$. The parameter $\delta$ therefore encodes the UV reflection phase of the charged scalar in the $\mathrm{AdS}_2$ cavity. The scalar cloud appears when this phase condition is compatible with regularity in the IR and with the chosen UV boundary condition.\newline
From the defect perspective, this is analogous to the supercritical regime of a charged Wilson line. The scalar-free defect saddle becomes unstable, and the charged scalar develops a screening cloud. In the full $\mathrm{AdS}_3$ problem studied in Sec.~\ref{sec:level4}, this instability is resolved by the formation of a fully backreacted hairy charged-BTZ branch.

\section{\label{sec:level6}Effective Field Description and Quantization}

We now develop an effective description of the charged scalar cloud in the near-horizon region. The relevant ingredients have already been derived in the previous sections: the near-horizon $\mathrm{AdS}_2$ geometry, the charged scalar equation, the electric BF threshold, the ingoing IR condition, and the UV mixed boundary condition. The purpose of this section is not to introduce a new gravitational solution, but rather to assemble these ingredients into an effective two-dimensional description of the cloud and its screening interpretation.
\subsection{Effective Field Regime and Background Geometry}

After reducing over the rigid circle of radius $\gamma$, a fixed angular harmonic of the charged scalar behaves as a complex scalar field $\Phi(t,r)$ propagating in the near-horizon $\mathrm{AdS}_2$ throat. We use a radial coordinate $r$ proportional to $1/x$, so that the $\mathrm{AdS}_2$ boundary lies at $r\to 0$, while the horizon lies at $r\to\infty$. In this coordinate the metric takes the Poincare form
\begin{equation} \label{Poincare-like coord.}
ds_{(2)}^2 = \frac{R_A^2}{r^2}
\left( -dt^2+dr^2 \right), \qquad R_A^2=\frac{\ell^2}{2}.
\end{equation}
In the original near-horizon coordinate $x$, we choose the gauge
\begin{equation}
A_t(x)=-Ex,
\end{equation}
so that $F_{tx}=E$. Equivalently, in the $r$-coordinate,
\begin{equation}\label{A_tANDF_rt}
A_t(r)=-\frac{E}{r},
\qquad
F_{rt}=\frac{E}{r^2}.
\end{equation}
Thus the coordinate component of the field strength is not constant in $r$, but the orthonormal electric field in the $\mathrm{AdS}_2$ throat is constant. For the extremal charged BTZ throat this physical electric field is
\begin{equation}
E=\frac{1}{2\sqrt{\pi G}\,\ell}.
\end{equation}
The effective two-dimensional theory contains a charged scalar coupled to this electric field. Writing
\begin{equation}
\Phi(t,r) = \frac{1}{\sqrt2} \,
\mathcal{R}(t,r)e^{i\pi(t,r)},
\end{equation}
the reduced action may be written, up to an overall normalization, as
\begin{equation}
\label{effective2D}
\begin{aligned}
S_{\rm eff} &=
\int d^2x\,\sqrt{-g_{(2)}}\,\Bigg[ -\frac{1}{2} (\partial\mathcal{R})^2
\\
&\qquad
-\frac12\mathcal{R}^2 (\partial_\mu\pi-qA_\mu)
(\partial^\mu\pi-qA^\mu)
\\
&\qquad -\frac12m_{\rm eff}^2\mathcal{R}^2 \\
&\qquad -\frac{1}{4g_2^2}F_{\mu\nu}F^{\mu\nu} \Bigg].
\end{aligned}
\end{equation}
Here
\begin{equation}
m_{\rm eff}^2 = m^2+\frac{L^2}{\gamma^2}
\end{equation}
is the effective mass of the fixed angular harmonic. For a stationary mode with frequency $\omega$, one may equivalently write
\begin{equation}
\Phi(t,r) = \frac{1}{\sqrt2} \, \mathcal{R}(r) e^{-i\omega t+i\pi(r)} .
\label{decomposition}
\end{equation}
The static cloud corresponds to the $\omega=0$ saddle. Choosing a constant phase gives a real scalar profile. This configuration carries no radial flux, although it can carry charge density through its coupling to $A_t$.

In the supercritical regime, the electric field drives the effective scalar mass below the $\mathrm{AdS}_2$ BF bound. Equivalently,
\begin{equation}
\mu=ib, \qquad b>0,
\end{equation}
and the two near-boundary falloffs become
\begin{equation}
R(r) = A_- r^{1/2-ib} + A_+ r^{1/2+ib} +\cdots .
\end{equation}
For a real stationary profile, the coefficients may be chosen so that $A_+=A_-^*$, giving the logarithmically oscillatory form
\begin{equation}
R(r) = \mathcal{C}\sqrt{r} \, \cos\left(
b\log\frac{r}{r_0} + \delta \right).
\label{scalarprofile-effective}
\end{equation}
The phase $\delta$ is fixed by the UV mixed boundary condition, or equivalently by the ratio $A_+/A_-$. The amplitude $\mathcal{C}$ is likewise determined by the UV boundary data and by the IR regularity condition.

Since $|\Phi|^2=\mathcal{R}^2/2$, the logarithmically oscillatory profile gives
\begin{equation}
|\Phi(r)|^2 = \frac{\mathcal{C}^2}{2} r
\cos^2\left( b\log\frac{r}{r_0} + \delta \right).
\end{equation}
After averaging over the rapid logarithmic oscillations, one obtains the coarse-grained envelope
\begin{equation}
\overline{|\Phi(r)|^2} =
\frac{\mathcal{C}^2}{4}r .
\label{coarse-grained-cloud}
\end{equation}
Thus the scalar density grows toward the horizon and vanishes toward the $\mathrm{AdS}_2$ boundary. This reflects the accumulation of low-energy charged modes in the redshifted near-horizon region. In the full $\mathrm{AdS}_3$ problem, this near-horizon cloud is resolved by the backreacted hairy charged-BTZ solutions constructed in Sec.~\ref{sec:level4}.\newline
The effective description also gives a simple way to visualize classical screening. The scalar condensate sources the electric field equation and modifies the electric flux along the throat. A representative simplified model is obtained by solving the coupled radial equations
\begin{equation}
r^2 R^{\prime\prime} - rR^\prime + A_t^2 r^2 R = 0,
\end{equation}
and
\begin{equation}
A_t^{\prime\prime} + \frac{2}{r}A_t^\prime = \kappa \frac{A_t R^2}{r^2},
\end{equation}
where $\kappa$ is an effective coupling controlling the strength of the scalar contribution to the electric field. These equations are not the full Einstein-Maxwell-scalar system. Rather, they define an effective throat model illustrating how a charged scalar cloud can screen the electric field.
\begin{figure*}[t]
\centering
\includegraphics[width=1.0\textwidth]{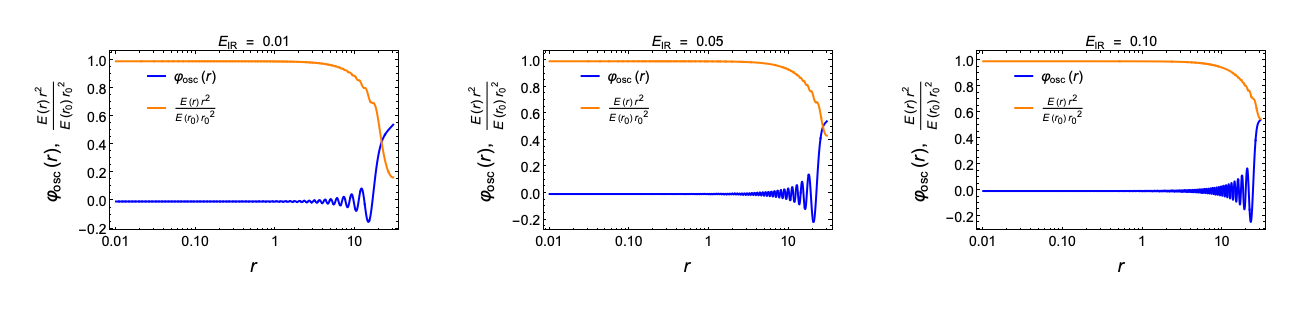}
\caption{Classical screening profiles in the effective $\mathrm{AdS}_2$ throat model. The blue curve shows the scalar profile $\phi(r)=\sqrt r\,R(r)$, while the orange curve shows the normalized electric-flux variable
$\mathcal F(r)/\mathcal F(r_{\rm UV})$, with
$\mathcal F(r)=r^2\mathcal E(r)$. The three panels correspond to increasing values of the imposed IR electric-field parameter,
$E_{\rm IR}=0.01,0.05,0.10$, from left to right. Increasing $E_{\rm IR}$ strengthens the scalar response and produces stronger depletion of the electric flux toward the UV, illustrating the screening-cloud interpretation in the effective throat description. These profiles are obtained from the simplified effective throat model and should not be confused with the full backreacted charged-BTZ solutions of Sec.~\ref{sec:level4}.}
\label{fig:effective-screening-profiles}
\end{figure*}
The profiles in Fig.~\ref{fig:effective-screening-profiles} should be understood as an effective-throat illustration rather than as the full backreacted charged-BTZ solution. They show that once the BF-violating scalar mode is excited, the scalar response is localized toward the near-horizon region and sources a radial variation of the electric flux. This supports the interpretation of the supercritical cloud as a screening configuration in the $\mathrm{AdS}_2$ throat. The fully backreacted realization of this mechanism in the complete $\mathrm{AdS}_3$ geometry was constructed numerically in Sec.~\ref{sec:level4}.

\subsection{Effective Screening Interpretation}

The screening interpretation can also be phrased in amplitude-phase variables. Writing
\begin{equation}
\Phi(t,r) =
\frac{1}{\sqrt2} \, \mathcal R(t,r)\, e^{i\pi(t,r)},
\qquad
\mathcal{R} \geq 0 ,
\end{equation}
the gauge-invariant combination is
\begin{equation}
B_\mu=\partial_\mu\pi-qA_\mu .
\end{equation}
The effective action contains the Stückelberg term
\begin{equation}
-\frac12\mathcal R^2 B_\mu B^\mu ,
\end{equation}
so a nonzero scalar amplitude gives the gauge field an effective mass in the throat. This should be understood in the semiclassical bulk sense: the local gauge symmetry is not broken, since the phase can be removed by a gauge choice.\newline
For the static cloud branch we may work in the zero-radial-flux sector, where the scalar phase is constant. The scalar then carries no radial current, but it still carries charge density through its coupling to $A_t$. Consequently the Maxwell equation is sourced by the condensate, and the electric flux is no longer constant along the throat. This is the effective-field-theory version of the screening mechanism: the trivial branch $\mathcal R=0$ describes the unscreened charged throat, while the nontrivial branch $\mathcal{R} \neq 0 $ describes a charged scalar cloud that partially screens the defect electric field.\newline
The role of the electric field is already visible in the scalar equation. With the convention used above \eqref{A_tANDF_rt}, the gauge coupling contributes a term proportional to $-q^2A_t^2\mathcal R$. Thus the electric field shifts the effective $\mathrm{AdS}_2$ mass downward. When the shift is large enough, the BF bound is violated, $\mu=ib$, and the scalar develops the logarithmically oscillatory cloud. The profiles shown in Fig.~\ref{fig:effective-screening-profiles} illustrate this mechanism in a simplified effective throat model. They should not be confused with the full backreacted charged-BTZ solutions of Sec.~\ref{sec:level4}; rather, they provide a local $\mathrm{AdS}_2$ interpretation of the same physical instability.

\section{\label{seclev:7}Zero-Mode Quantization and Cloud Charge}

We now discuss the quantization of the lowest gauge--phase degree of freedom associated with the scalar cloud. This construction should be understood as an effective quantization of the collective mode localized in the near-horizon $\mathrm{AdS}_2$ throat, following the logic of Appendix~A.4 of Ref.~\cite{Aharony:2023amq}, adapted to the present geometry. The goal is not to rederive the full backreacted charged-BTZ solution, but to identify the charge carried by the low-energy cloud mode in the effective throat description.\newline
We expand around a static scalar-cloud background in the effective $\mathrm{AdS}_2$ theory. The background profiles $R_s(r)$ and $A_t^{(s)}(r)$ should be viewed as the near-horizon, coarse-grained description of the cloud branch constructed in Sec.~\ref{sec:level4}, or equivalently as a static background of the effective throat model discussed in Sec.~\ref{sec:level6}. They are not the scalar-free charged-BTZ background of Sec.~\ref{sec:level2}. We work in radial gauge,
\begin{equation}
A_r^{(s)}=0 .
\end{equation}
We parameterize the scalar as
\begin{equation}
\Phi(t,r) =
\frac{1}{\sqrt2} \left[ R_s(r)+\sigma(t,r) \right]
e^{i\vartheta(t,r)},
\end{equation}
and expand the gauge field as
\begin{equation}
A_\mu(t,r) = A_\mu^{(s)}(r)+a_\mu(t,r).
\end{equation}
At linear order, the gauge transformations act as
\begin{equation}
\vartheta \to \vartheta + q\alpha,
\qquad a_\mu\to a_\mu+\partial_\mu\alpha,
\qquad
\sigma \to \sigma .
\end{equation}
Thus $\sigma$ is gauge invariant, while $\vartheta$ is the phase fluctuation. The gauge-invariant combination in the phase sector is
\begin{equation}
\partial_\mu\vartheta-q a_\mu .
\end{equation}
The gauge-invariant combination in the phase sector is
\begin{equation}
B_\mu \equiv
\partial_\mu\vartheta-q a_\mu .
\end{equation}
Keeping only the phase-gauge sector, the quadratic action becomes
\begin{equation}
\label{eq:Squad-full}
\begin{aligned}
S_{\rm quad}[\vartheta,a_\mu] =&
\int dt\,dr\, \sqrt{-g_{(2)}}\,
\Bigg[ -\frac12 R_s(r)^2 B_\mu B^\mu
\\
&\hspace{1.7cm} -\frac{1}{4g_2^2} F_{\mu\nu}(a)F^{\mu\nu}(a) +\cdots
\Bigg].
\end{aligned}
\end{equation}
where the ellipsis denotes the radial amplitude fluctuation $\sigma$ and other massive modes. The low-energy structure is therefore controlled by the Stückelberg coupling between the scalar phase and the gauge field.
We impose radial gauge,
\begin{equation}
a_r(t,r)=0 .
\end{equation}
Residual gauge transformations preserving this condition satisfy $\partial_r\alpha=0$, so $\alpha=\alpha(t)$. In the static sector, the $a_t$ equation obtained from \eqref{eq:Squad-full} is the Proca--Gauss equation
\begin{equation}
\partial_r
\left( \frac{r^2}{R_A^2 g_2^2} \partial_r a_t
\right) - q^2 R_s(r)^2 a_t = 0 .
\label{eq:at-eq-general}
\end{equation}
For a generic cloud profile, this equation must be treated with the full $r$-dependent function $R_s(r)^2$. However, in the supercritical regime we may use the coarse-grained envelope of the logarithmically oscillatory cloud,
\begin{equation}
R_s(r)^2
\simeq \frac{\mathcal C^2}{2} \, r .
\label{eq:Rs-envelope}
\end{equation}
Substituting \eqref{eq:Rs-envelope} into \eqref{eq:at-eq-general} gives
\begin{equation}
r^2a_t''(r) + 2r a_t'(r) - \kappa r\,a_t(r) = 0, \qquad
\kappa = \frac{q^2g_2^2R_A^2\mathcal C^2}{2}.
\label{eq:at-ode-kappa}
\end{equation}
The constancy of $\kappa$ is a consequence of the coarse-grained scaling $R_s^2\propto r$. Without this approximation, one should keep the full equation \eqref{eq:at-eq-general}. Equation \eqref{eq:at-ode-kappa} has the Bessel-function solution
\begin{equation}
a_t(r) = \frac{C_1}{\sqrt{\kappa r}}\,
I_1 \! \left(2\sqrt{\kappa r}\right)
+ \frac{2C_2}{\sqrt{\kappa r}} \,
K_1\!\left(2\sqrt{\kappa r}\right).
\label{eq:at-bessel}
\end{equation}
Near the $\mathrm{AdS}_2$ boundary, $r\to0$, the two independent behaviors are
\begin{equation}
\begin{split}
a_t^{(1)}(r) &\sim \mathrm{const.} +\mathcal O(r),
\\ a_t^{(2)}(r)
& \sim \frac{1}{\kappa r} + 2\left( \log\sqrt{\kappa r}
+\gamma_E -\frac{1}{2} \right) +\cdots .
\end{split}
\label{eq:at-UV}
\end{equation}
The $1/r$ branch is the $\mathrm{AdS}_2$ analogue of a Coulombic electric potential near the defect.
The quantization of the cloud charge is controlled by the lowest phase--gauge collective mode. At very low energies we project onto
\begin{align}
\vartheta(t,r)
& = \hat{x}(t) +\cdots , \\
a_t(t,r)
& = a_t^{({\rm nor})}(r)\,\hat{p}(t) +\cdots ,
\label{eq:zm-ansatz}
\end{align}
where $a_t^{({\rm nor})}(r)$ is a fixed normalized static profile and the ellipsis denotes higher radial modes. The variable $\hat{x}$ is the collective phase of the cloud, while $\hat{p}$ is its conjugate charge variable after an appropriate normalization.\newline
The symplectic term comes from the time-derivative part of \eqref{eq:Squad-full}. Since
\begin{equation}
\sqrt{-g_{(2)}}\,g^{tt}=-1
\end{equation}
in the Poincare metric \eqref{Poincare-like coord.}, the relevant cross term is
\begin{equation}
S_{\rm quad} \supset -q \int dt\,dr\,
R_s(r)^2\,a_t(t,r)\,
\partial_t\vartheta(t,r).
\end{equation}
Using \eqref{eq:zm-ansatz}, this becomes
\begin{equation}
S_{\rm zm} \supset -q
\left[ \int_{r_0}^{r_{\rm IR}}dr\,
R_s(r)^2a_t^{({\rm nor})}(r)
\right] \int dt\,\hat{p}(t)\dot{\hat{x}}(t).
\label{eq:Szm-sympl}
\end{equation}
We choose the normalization of $a_t^{({\rm nor})}$ so that
\begin{equation}
q \int_{r_0}^{r_{\rm IR}}dr \, R_s(r)^2a_t^{({\rm nor})}(r) = 1,
\label{eq:norm-at}
\end{equation}
up to an overall sign that can be absorbed into the definition of $\hat{p}$. With this choice the zero-mode action contains the canonical term
\begin{equation}
S_{\rm zm} \supset \int dt\,\hat{p}\dot{\hat{x}},
\end{equation}
and therefore
\begin{equation}
[\hat{x},\hat{p}]=i .
\end{equation}
The coordinate $\hat{x}$ is a compact phase. Taking its period to be $2\pi$, the conjugate momentum is quantized,
\begin{equation}
|n\rangle=e^{in\hat{x}}|0\rangle,
\qquad n\in\mathbb Z,
\qquad \hat{p}|n\rangle=n|n\rangle .
\end{equation}
Thus the cloud charge is quantized in units determined by the normalization of the zero-mode profile. With the normalization \eqref{eq:norm-at}, the integer $n$ labels the charge carried by the cloud mode.\newline
If one keeps the normalization explicit, it is useful to define
\begin{equation}
q_{\rm eff} = q \int_{r_0}^{r_{\rm IR}}dr\,
R_s(r)^2a_t^{({\rm nor})}(r).
\end{equation}
Then the canonical momentum conjugate to the compact phase is $q_{\rm eff}\hat{p}$. Equivalently, the charge lattice of the cloud is fixed by $q_{\rm eff}$. In physical units one may write
\begin{equation}
Q_{\rm cloud}=nq_{\rm eff}, \qquad n\in\mathbb Z .
\end{equation}
Let $Q_{\rm UV}$ denote the electric charge measured at the UV cutoff. In the effective zero-mode description, the residual charge after the cloud forms is
\begin{equation}
Q_{\rm res} = Q_{\rm UV}-Q_{\rm cloud} =
Q_{\rm UV}-nq_{\rm eff}, \qquad n\in \mathbb{Z} .
\end{equation}
Complete screening is possible only if the UV charge lies on the cloud-charge lattice,
\begin{equation}
Q_{\rm UV}=nq_{\rm eff}.
\end{equation}
Otherwise the scalar cloud can reduce the electric flux but cannot cancel the charge completely. This is the sense in which the effective theory predicts partial screening for a generic charge.\newline
A gauge-invariant diagnostic of this screening is the radial electric flux
\begin{equation}
\mathcal E(r)
\equiv
\frac{r^2}{R_A^2g_2^2}A_t'(r).
\label{eq:flux-def}
\end{equation}
Gauss' law relates the radial variation of $\mathcal E(r)$ to the charge density stored in the cloud. Therefore the difference between the flux at the UV cutoff and the flux near the IR end of the throat measures the integrated charge carried by the scalar cloud:
\begin{equation}
Q_{\rm cloud} = \mathcal E(r_{\rm UV}) - \mathcal E(r_{\rm IR}) .
\end{equation}
This completes the effective quantization picture. The scalar cloud supports a compact collective phase, its conjugate momentum is quantized, and the corresponding charge can partially screen the electric flux sourced by the defect.

\begin{figure}[t]
\centering
\includegraphics[width=\columnwidth]{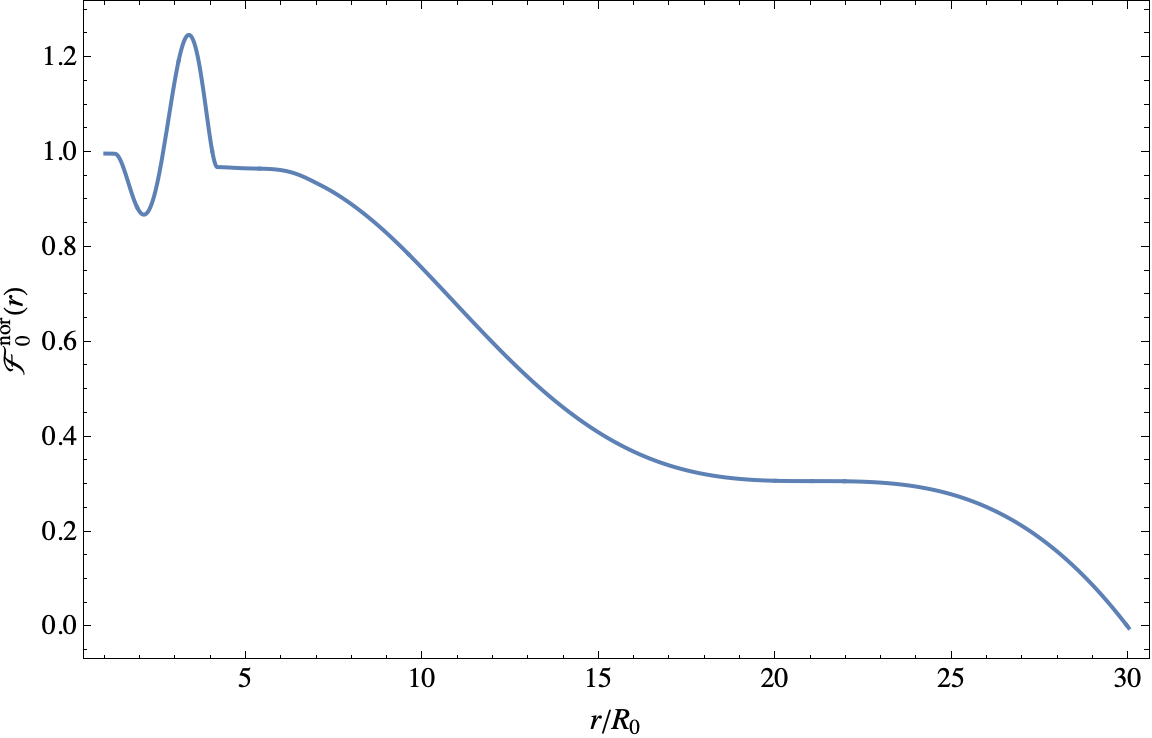}

\vspace{2mm}

\includegraphics[width=\columnwidth]{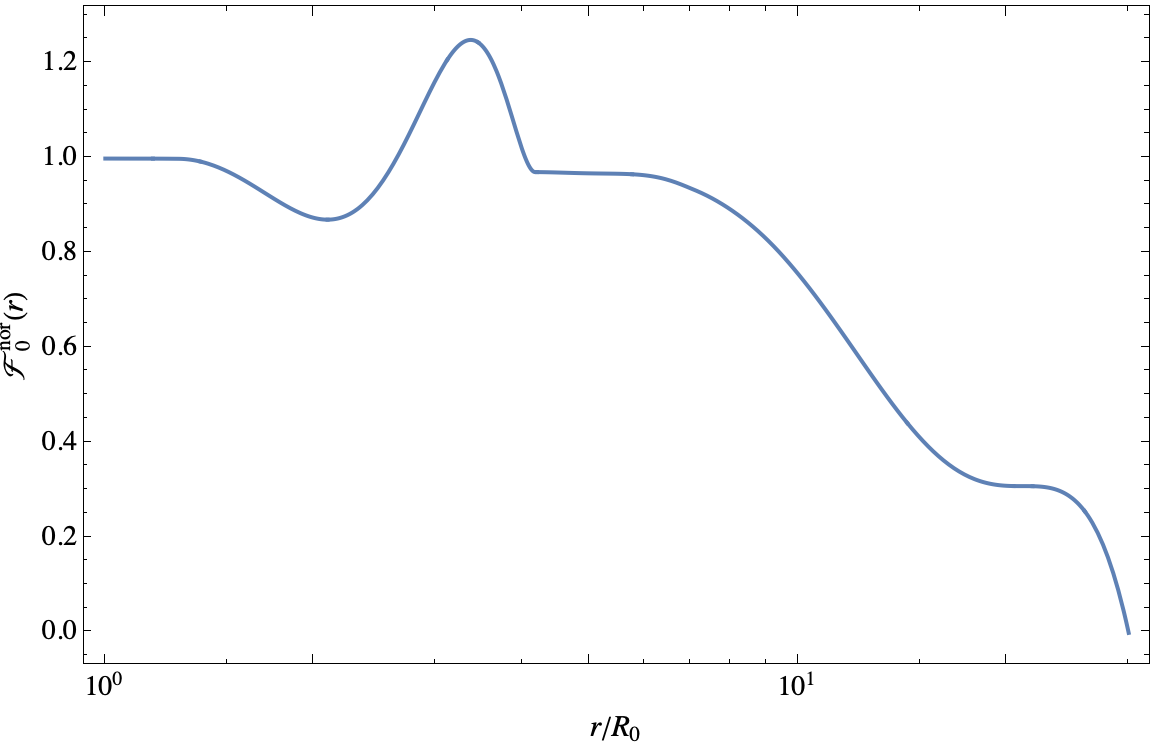}

\caption{
Representative normalized zero-mode electric-flux profile in the effective $\mathrm{AdS}_2$ throat model. The plotted quantity is the normalized flux mode $\mathcal F_0^{\rm nor}(r)$ associated with the static gauge fluctuation $a_t^{({\rm nor})}(r)$. The profile is obtained by solving the static Proca-Gauss equation using an illustrative logarithmically oscillatory scalar-cloud background. It should therefore be viewed as a visualization of the effective zero-mode construction, not as an additional numerical solution of the full backreacted charged-BTZ system.}
\label{fig:flux_profiles}
\end{figure}

\section{Discussion and Outlook}
We have studied the instability of a charged scalar in the background of an extremal charged BTZ black hole. The instability is triggered by the violation of the effective AdS$_2$ Breitenlohner--Freedman bound in the near-horizon region. The fixed-background analysis determines the onset of the instability, while the nonlinear analysis shows that the unstable mode continues into a finite-temperature branch of fully backreacted, source-free, node-free hairy solutions.\newline
The backreacted solutions confirm that a finite scalar condensate cannot be consistently described on an unchanged charged-BTZ background. Both the Maxwell field and the metric respond to the scalar profile. The critical ratio $T_c/Q_c$ decreases as the effective backreaction parameter is increased, and the condensate amplitude is correspondingly suppressed. Near the critical point, the condensate exhibits the mean-field scaling behavior
$\psi_+/Q^{\Delta_+}\propto[1-(T/Q)/(T_c/Q_c)]^{1/2}$.\newline
The numerical solutions constructed here are restricted to the near-critical finite-temperature regime. They therefore establish the nonlinear continuation of the instability, but do not determine its exact zero-temperature endpoint. In particular, the finite-temperature solutions do not possess an exact AdS$_2\times S^1$ throat. A separate extremal near-horizon analysis indicates that the original regular AdS$_2\times S^1$ geometry cannot remain unchanged in the BF-violating condensed phase within the power-law ansatz considered here. The infrared geometry or electric-field profile must instead reorganize.\newline
The effective AdS$_2$ description developed in the later sections provides a complementary interpretation in terms of electric-flux screening. The scalar response redistributes the electric flux and may be viewed as the gravitational analogue of a screening cloud around a charged impurity or line defect. This interpretation is consistent with the fully backreacted charged-BTZ solutions, although the effective throat model should not be identified with the complete nonlinear geometry.\newline
Several questions remain open. The most direct extension is to continue the nonlinear branch toward $T/Q\to0$ and determine the resulting extremal infrared geometry. It would also be useful to compare the thermodynamic potentials of the hairy and normal charged-BTZ branches and establish which phase is dominant in the relevant ensemble. Further directions include time-dependent evolution of the instability, higher-dimensional generalizations, and a sharper relation between the bulk screening mechanism and the charged operator spectrum of the dual field theory.

\acknowledgments
 I thank Zohar Komargodski for pointing out the original connection
 that motivated this work, for many insightful discussions and extensive correspondence, and for carefully reading an early version of the manuscript.\newline
 I also would like to thank Mark Mezei for useful discussions about the BF-bound in $\rm AdS_d$ and also Borut Bajc for many useful conversations and discussions during this work. Finally, I thank the referee for constructive comments, particularly regarding the gravitational backreaction analysis, which helped improve the manuscript.

\appendix

\section{UV and near-horizon expansion}

\subsection{UV expansion}
We need to know how fields look when $\rho \to \infty$, specifically $f(\rho)$ and $e^{-\chi(\rho)} f(\rho)$, therefore we want \emph{large-$\rho$} expansion of the fields. Therefore we have
\begin{equation}
    f(\rho) \sim \frac{\rho^2}{\ell^2} - 8 \pi G^2 \log{\frac{\rho}{\ell}}- 8 GM + \dots
\end{equation}
and 
\begin{equation}
    A_t(\rho) \sim Q \log{\frac{\rho}{\ell}} + \mu  + \dots,
\end{equation}
where compared to \cite{Hartnoll:2008vx,Hartnoll:2008kx}, now the gauge field has a logarithmic term and the constant $M$ is the BTZ mass parameter after substracting the logarithmic electric contribution.\newline
At large $\rho$, 
\begin{equation}
    f(\rho) \sim \frac{\rho^2}{\ell^2}, \quad \chi(\rho)  \to 0,
\end{equation}
and the gauge field term is subleading because 
\begin{equation}
    \frac{A_t^2}{f^2} \sim \frac{\log^2{\rho}}{\rho^4}
\end{equation}
so the scalar equation now reduces to
\begin{equation}
    \psi^{\prime \prime} +\frac{3}{\rho} \psi^\prime - \frac{m^2 \ell^2}{\rho^2} \psi =0. 
\end{equation}
For $\psi \sim \rho^{-\Delta} $, we get
\begin{equation}
    \Delta_{\pm} = 1 \pm \sqrt{1+ m^2 \ell^2}
\end{equation}
Therefore, the scalar field has an UV expansion
\begin{equation}
    \psi = \psi_- \rho^{-\Delta_-} + \psi_+ \rho^{-\Delta_+}+ \dots. 
\end{equation}
One thing to keep in mind is that this is the $\rm AdS_3$ falloff and it is not the same as the $\rm AdS_2$ throat falloff from \ref{sec:level3}. \newline
On the other hand, Maxwell equation is 
\begin{equation}
    A_t^{\prime \prime} +\left( \frac{1}{\rho} + \frac{\chi^2}{2} \right)A_t^\prime - \frac{q^2 \psi^2}{f}A_t= 0,
\end{equation}
and at leading order, since $\psi \to 0$ and $\chi^\prime \to 0$ this becomes
\begin{equation}
    A_t^{\prime \prime} + \frac{1}{\rho} A_t^\prime = 0
\end{equation}
giving us
\begin{equation}
    A_t(\rho) = Q \log{\frac{\rho}{\ell}} + \mu + \text{(subleading scalar corrections)}. 
\end{equation}
Schematically, scalar current starts as
\begin{equation}
    \delta A_t = \mathcal{O}\left( \rho^{-2 \Delta_*} \log{\rho} \right)
\end{equation}
where $\Delta_*$ is the leading allowed scalar falloff. For example, if we impose the standard no-source condition $\psi_-=0$ then $\Delta_*=\Delta_-$. Therefore, $Q$ is the coefficient of $\log{\rho}$ while $\mu$ is the consant part after choosing a renormalization sccale, which is scheme dependent due to logarithm. \newline
Using Einstein's equations for $f$ we have
\begin{align} \label{fprime}
    f^\prime  &= \frac{2 \rho}{\ell^2} \nonumber \\
   & - 8 \pi G \rho \left[ e^{-\chi(\rho)} \left( A_t^\prime\right)^2 + f \left( \psi^\prime \right)^2 + \frac{q^2 e^{\chi} A_t^2}{f} \psi^2 +m^2 \psi^2 \right] =0
\end{align}
At leading order, $A_t^\prime= \frac{Q}{\rho}$, so the gauge contribution gives
\begin{equation}
    -8 \pi G\rho \left( A_t^\prime \right)^2 = - \frac{8 \pi G Q^2}{\rho}
\end{equation}
Integrating \eqref{fprime} we have
\begin{equation}
    f(\rho)  = \frac{\rho^2}{\ell^2}- 8 \pi G Q^2 \log{\frac{\rho}{\ell}}-8 GM + \dots 
\end{equation}
and for $\psi(\rho) \sim a \rho^{-\Delta}$ we obtain
\begin{equation}
    f \left( \psi^\prime \right)^2 + m^2 \psi^2 \sim \frac{2 \Delta(\Delta-1)}{\ell^2} a^2 \rho^{-2\Delta}
\end{equation}
and the scalar contribution to $f$ is 
\begin{equation}
    \delta f_\psi=\frac{8 \pi G \Delta}{\ell^2}a \rho^{2-2\Delta}+ \dots .
\end{equation}
Therefore, for a single leading scalar branch
\begin{equation}
    f(\rho) = \frac{\rho^2}{\ell^2} -8 \pi G Q^2 \log{\frac{\rho}{\ell}}- 8 G M + \frac{8 \pi G \Delta }{\ell^2}a^2 \rho^{2-2 \Delta}+ \dots .
\end{equation}
For the metric expression $\chi(\rho)$, using
\begin{equation}
    \chi^\prime =-16 \pi G \rho \left[ \left( \psi^\prime \right)^2 + \frac{q^2 e^{\chi} A_t^2}{f} \psi^2 \right]
\end{equation}
the second term is more suppressed so, at leading order we have
\begin{equation}
    \chi^\prime = -16 \pi G \rho \left( \psi^\prime \right)^2
\end{equation}
and for $\psi \sim a \rho^{-\Delta}$ we obtain
\begin{equation}
    \chi(\rho) = \chi_\infty + 8 \pi G \Delta a^2 \rho^{-2\Delta}+  \dots 
\end{equation}
and choosing the boundary time normalization $\chi_\infty=0$ we get
\begin{equation}
    \chi(\rho) = 8 \pi G \Delta a^2 \rho^{-2\Delta}+ \dots .
\end{equation}
Finally, using $e^{-\chi} f \approx (1-\chi)f$, we have
\begin{equation}
    e^{-\chi} f \approx \frac{\rho^2}{\ell^2}- 8 \pi G Q^2 \log{\frac{\rho^2}{\ell^2}}- 8 G M + \dots. 
\end{equation}
So $f$ itself can contain scalar dependent UV piece, but the physica $g_{tt}$ component (combination) $e^{-\chi} f$ gives a mass read off. We read $M$ from the constnat term after substracting $\rm AdS$ piece of Maxwell logarithm. Therefore, when we solve numerically form horizon outward, we will read off $Q, \mu, M, \psi_-, \psi_+$. 

\subsection{Near-horizon expansion}

We now consider the near-horizon expansion around a nonextremal horizon
\begin{equation}
    \rho=\rho_h ,
\end{equation}
where
\begin{equation}
    f(\rho_h)=0 .
\end{equation}
Regularity of the gauge field at the future horizon requires working in a gauge where
\begin{equation}
    A_t(\rho_h)=0 .
\end{equation}
We therefore expand the fields as
\begin{align}
    f(\rho) &= f_1(\rho-\rho_h)+f_2(\rho-\rho_h)^2+\cdots , \\
    \chi(\rho) &= \chi_h+\chi_1(\rho-\rho_h)+\cdots , \\
    A_t(\rho) &= E_h(\rho-\rho_h)+A_2(\rho-\rho_h)^2+\cdots , \\
    \psi(\rho) &= \psi_h+\psi_1(\rho-\rho_h)+\psi_2(\rho-\rho_h)^2+\cdots .
\end{align}
Here $E_h=A_t'(\rho_h)$ is the electric field at the horizon, while $\psi_h$ and $\chi_h$ are the horizon values of the scalar and redshift function. These quantities serve as shooting data for the numerical construction. \newline
The equations of motion determine the higher coefficients in terms of the horizon data. From the Einstein equation for $f$, one obtains
\begin{equation}
f_1 = \frac{2\rho_h}{\ell^2}- 8\pi G\,\rho_h
\left[ e^{-\chi_h}E_h^2 + m^2\psi_h^2 \right] .
\end{equation}
The scalar equation contains a potentially singular term proportional to $1/f$ near the horizon. Requiring regularity fixes the first scalar coefficient as
\begin{equation}
    \psi_1 = \frac{m^2}{f_1}\psi_h .
\end{equation}
Similarly, the Maxwell equation fixes the next coefficient in the gauge-field expansion. Writing
\begin{equation}
    A_t(\rho)=E_h(\rho-\rho_h)+A_2(\rho-\rho_h)^2+\cdots ,
\end{equation}
one finds
\begin{equation}
A_2 = \frac{E_h}{2}
\left[ \frac{q^2\psi_h^2}{f_1} - \frac{1}{\rho_h} -
\frac{\chi_1}{2}
\right] .
\end{equation}
The first coefficient in the expansion of $\chi$ is obtained from the $\chi$ equation,
\begin{equation}
\chi_1=-16\pi G\,\rho_h
\left[ \psi_1^2 + \frac{q^2 e^{\chi_h}E_h^2}{f_1^2}\psi_h^2
\right] .
\end{equation}
Thus, once the four horizon parameters
\begin{equation}
    \{\rho_h,\psi_h,E_h,\chi_h\}
\end{equation}
are specified, the near-horizon expansion is fixed order by order.\newline
For numerical integration, one starts slightly outside the horizon,
\begin{equation}
    \rho=\rho_h+\epsilon ,
    \qquad
    0<\epsilon\ll 1 ,
\end{equation}
using the Taylor-expanded fields
\begin{align}
    f(\rho_h+\epsilon) &= f_1\epsilon+\mathcal{O}(\epsilon^2), \\
    \chi(\rho_h+\epsilon) &= \chi_h+\chi_1\epsilon+\mathcal{O}(\epsilon^2), \\
    A_t(\rho_h+\epsilon) &= E_h\epsilon+\mathcal{O}(\epsilon^2), \\
    \psi(\rho_h+\epsilon) &= \psi_h+\psi_1\epsilon+\mathcal{O}(\epsilon^2).
\end{align}
The shooting parameters are then adjusted so that the UV boundary conditions are satisfied, in particular the desired mixed or no-source condition on the scalar field.\newline
The Hawking temperature associated with the horizon is
\begin{equation}
    T= \frac{e^{-\chi_h/2}f_1}{4\pi} .
\end{equation}
If the numerical solution is obtained with $\chi(\infty)=\chi_\infty\neq0$, one must finally rescale the time coordinate in order to impose the standard boundary normalization. The physical temperature is then
\begin{equation}
    T_{\rm phys}= e^{\chi_\infty/2} \frac{e^{-\chi_h/2}f_1}{4\pi} =
    \frac{e^{-(\chi_h-\chi_\infty)/2}f_1}{4\pi}.
\end{equation}
In the gauge where $\chi_\infty=0$, this reduces to the expression above.

\section{Variation of action}
\subsection{Variation of the bulk action}
Variation of the bulk term is given by:
\begin{equation} \label{bulkvariation}
    \delta S_{ \rm bulk} = \frac{1}{2} \int_{r=r_0} dt \left[ \Phi^* \partial_r \delta \Phi + (\partial \Phi)^* \delta \Phi \right]
\end{equation}
where $\Phi(r)$ and $\partial_r \Phi(r)$ are given by:
\begin{align}
    \Phi(r) & \simeq A_- r^{\Delta_-} +A_+ r^{\Delta_+} \\
    \partial_r\Phi & \simeq \Delta_- A_- r^{\Delta_- -1}+ \Delta_+ A_+ r^{\Delta_+ -1}
\end{align}
and the variation of the two above equations is given by:
\begin{align}
    \delta \Phi & \simeq \delta A_-r^{\Delta_-} + \delta A_+r^{\Delta_+} \\
    \partial (\delta \Phi) & \simeq \Delta_- \delta A_-r^{\Delta_--1} +\Delta_+\delta A_+ r^{\Delta_+ -1}.
\end{align}
Expading the first term of \eqref{bulkvariation}, we have
\begin{align}
    \Phi^*\partial_r \delta\Phi \simeq & \left( A_-^* r^{\Delta_-} + A_+^* r^{\Delta_+} \right) \\
    &\left( \Delta_- \delta A_-r^{\Delta_- -1} + \Delta_+ \delta A_+r^{\Delta_+ -1} \right),
\end{align}
which gives four terms, but what matters to us is the leading term 
\begin{equation}
   \Phi^*\partial_r \delta\Phi \supset A_-^* \Delta_- \delta A_- r^{2\Delta-1}.
\end{equation}
Using $\Delta_- = 1/2 - \mu$, we see that $2\Delta_- - 1 =-2\mu $, therfore for first term in \eqref{bulkvariation} variation we can write:
\begin{equation} \label{firsterm}
    \Phi^*\partial_r \delta\Phi \supset \Delta_- r^{-2 \mu} A_-^*  \delta A_-.
\end{equation}
Similaryl, for the second term we have:
\begin{equation}
\left( \partial_r \Phi  \right)^* \delta \Phi \simeq \left(\Delta_- A_-^* r^{\Delta_- -1 } + \dots \right) \left( \delta A_- r^{\Delta_-} +\dots  \right),
\end{equation}
and again, the relevant terms for us are 
\begin{equation} \label{secondterm}
  \left( \partial_r \Phi  \right)^* \delta \Phi  \supset\Delta_- r^{-2 \mu} A_-^* \delta A_-.
\end{equation}
Now we combine \eqref{firsterm} and \eqref{secondterm} and substitute them into \eqref{bulkvariation} at the UV cutoff $r=r_0$ to obtain the following:
\begin{equation}
    \delta S_{ \rm bulk} =\frac{1}{2} \int dt \left[\Delta_- r_0^{-2 \mu} +  \Delta_- r_0^{-2 \mu} \right]A_-^* \delta A_- + \text{c.c}.
\end{equation}
Finally, using $\Delta_-= 1/2- \mu$, we obtain equation:
\begin{align}
    \delta S_{ \rm bulk} = \int dt \left[ \frac{1-2\mu}{2}r_0^{-2 \mu} A_-^* \delta A_- \right] + \text{c.c}.
\end{align}

\subsection{Variation of the first boundary term}
We now add a local boundary term at the cutoff $r=r_0$,
\begin{equation}
    S_{\rm bdy}^{(1)} = - \frac{1-2 \mu}{2} \int_{r=r_0} dt \sqrt{-\hat g} \mid \Phi \mid^2. 
\end{equation}
Near the boundary $\mid \Phi \mid^2$ is now given by:
\begin{equation}
\begin{split}
    \mid \Phi \mid^2 \simeq \mid & A_- \mid^2 r^{-2 \Delta_-} + \left( A_-^* A_+ + A_+^* A_- \right) r^{\Delta_- + \Delta_+ } \\
    &+\mid A_+ \mid^2 r^{-2 \Delta_+}
\end{split}
\end{equation}

where the leading term now is given by:
\begin{equation} \label{nearbdy-Phisol.}
    \mid \Phi \mid^2 \supset \mid A_- \mid^2 r^{1-2 \mu} +\dots
\end{equation}
where $2 \Delta_-= 1-2\mu$ and the induced metric on $r=r_0$ gives us a factor of $\sqrt{ \hat g} \propto r_0^{-1}$. 
Taking its variation we find the following:
\begin{equation}
    \delta S_{\rm bdy}^{(1)} = - \frac{1-2 \mu}{2} \int_{r=r_0} dt \sqrt{-\hat g} \left( \Phi^* \delta \Phi + \Phi \delta \Phi \right).
\end{equation}
We can now write:
\begin{equation}
\begin{aligned}
\delta S_{\rm bdy}^{(1)}
= -\,\frac{1-2\mu}{2}
\int d\omega \,\Big[
& C\, r_0^{-2\mu}
  \big(A_-^* \delta A_- + \text{c.c.}\big)  \\
& + C
  \big(A_+^* \delta A_- 
      + A_-^* \delta A_+ 
      + \text{c.c.}\big)  \\
& + C\, r_0^{2\mu}
  \big(A_+^* \delta A_+ + \text{c.c.}\big)
\Big].
\end{aligned}
\end{equation}
\subsection{Variation of the second term}
We now perturb this $\rm DCFT_1$ by the relevant $\mid \Phi \mid^2$. In $\rm AdS$ language this is a double-trace deformation. Therefore, we add the following second boundary term
\begin{equation}
    S_{\rm bdy}^{(2)} =  -f_0 \int_{r=r_0} dt \sqrt{- \hat g} r_0^{2\mu} \mid \Phi \mid^2. 
\end{equation}
Near the boundary our solution is given by \eqref{nearbdy-Phisol.}, so 
\begin{equation}
    \mid \Phi \mid^2 \supset \mid A_- \mid^2 r^{1-2 \mu} + \dots,
\end{equation}
and from our induced metric, $\sqrt{- \hat g} \sim R_A/ r$, we find that 
\begin{equation}
    \sqrt{- \hat g} \mid \Phi \mid^2 \supset C \mid A_- \mid^2 r^{-2 \mu}.
\end{equation}
Multiplying by $r_0^{2 \mu}$ makes $\mid A_- \mid^2$ contribution finite at the cutoff:
\begin{equation}
    \sqrt{- \hat g} r_{0}^{2\mu} \mid \Phi \mid^2 \supset C \mid A_- \mid^2 + \mathcal{O} \left( r_0^{2 \mu} \right).
\end{equation}
So, by adding the second boundary term, we are re-introducing an $A_-^* A_-$ term with a finite coefficient $f_0$, which will compete with the cross term from $S_{\rm bulk} +S_{\rm bdy}^{(1)}$. \newline
At leading order in small $r_0$ the variation of the second boundary term is given by:
\begin{equation}
    \delta S_{\rm bdy}^{(2)} \simeq -f_0 C \int d\omega \left( A_-^* \delta A_- + A_- \delta A_-^*\right) + \mathcal{O} \left( r_0^{2 \mu} \right).
\end{equation}
After adding both boundary terms, the leading parts of the variation are:
\begin{itemize}
    \item from $S_{\rm bulk} + S_{\rm bdy}^{(1)}$: the cross term 
    \begin{equation}
        \sim 2 \mu \left( A_+^* \delta A_- + A_-^* \delta A_+ \right)
    \end{equation} \\
    \item from $\delta S_{\rm bdy}^{(2)}$ we have 
    \begin{equation}
        \sim f_0 \left( A_-^* \delta A_- + \text{c.c} \right)
    \end{equation}
\end{itemize}
So at leading order in small $r_0$ for the full variation we have
\begin{equation}
\begin{aligned}
\delta\!\left(S_{\rm bulk}+S_{\rm bdy}^{(1)}+S_{\rm bdy}^{(2)}\right)
\simeq \int d\omega \Big[
     2\mu\!\left(A_+^{*}\,\delta A_- + A_-^{*}\,\delta A_+\right)
\\
\qquad\qquad\qquad\quad
     +\, f_0\!\left(A_-^{*}\,\delta A_- + A_-\,\delta A_-^{*}\right)
\Big].
\end{aligned}
\end{equation}

\subsection{Derivation of Beta-function}
We start with our boundary condition
\begin{equation}
    \frac{A_+}{A_-} = R(r_0,f) =  -\frac{f}{r_0^{2\mu} \left[f-2\mu \right]}.
\end{equation}
This ratio is a physical quantity and it depends on $r_0$, therefore we write
\begin{equation} \label{b.c.1}
    0=r_0 \frac{d}{d r_0} R(r_0, f(r_0)) =  \left( r_0 \partial_{r_0} - -\beta_f \partial_f \right) R(r_0, f(r_0))
\end{equation}
For fixed $f$  we find that 
\begin{equation}
  r_0  \partial_{r_0} R(r_0, f) = \frac{2 \mu f}{f-2\mu} r_0^{-2\mu}.
\end{equation}
Now holding $r_0$ fixed, varying $f$ we have:
\begin{equation}
    R(r_0, f)= - \frac{1}{r_0^{2\mu}} \frac{f}{f-2\mu} = - \frac{1}{r_0^{2\mu}} G(f),
\end{equation}
where the derivative of $G(f)$ with respect to $f$ is given by
\begin{equation}
    \partial_f G(f) =  -\frac{2 \mu}{\left( f-2 \mu\right)^2}.
\end{equation}
Then we can write
\begin{equation}
    \partial_f R(r_0, f) = \frac{2 \mu}{ r_0^{2\mu} \left( f-2 \mu \right)^2}
\end{equation}
Finally for $\eqref{b.c.1}$ we have
\begin{align}
     0 &=r_0 \partial_{r_0} R-\beta_f \partial_f R \\
     &= \frac{2 \mu f}{f-2\mu} -\beta_f \frac{2 \mu}{ r_0^{2\mu} \left( f-2 \mu \right)^2} \\
     & = \frac{2 \mu}{ r_{0}^{2 \mu} \left( f-2 \mu \right)^2} \left[ f(f-2\mu ) -\beta_f \right],
\end{align}
from which we get our beta-function
\begin{equation}
    \beta_f=  -2\mu f +f^2. 
\end{equation}

\section{ Boundary Two-Point Function and Double-Trace Flow in the AdS\texorpdfstring{$_2$}{2} Throat}

In this section we compute the Euclidean propagator for the boundary mode dual to the  radial solution of the charged scalar field in the near-horizon $\mathrm{AdS}_2$ region of the 
near-extremal charged BTZ black hole.  We work entirely with the Whittaker solution obtained 
in the main text, imposing IR infalling boundary conditions and extracting the near-boundary 
coefficients $A_{\pm}(\omega)$.  These coefficients determine the Euclidean Green's function
$G^{(0)}(\omega)$ at the alternate quantization fixed point.  We then add a double-trace 
boundary deformation and obtain the exact resummed propagator.

\subsection{ Bulk Setup and Near-Boundary Expansion}
In the $\mathrm{AdS}_2$ throat we use the Poincar\'e metric
\begin{equation}
    ds^2 = \frac{dr^2 + dt_E^2}{r^2},
\end{equation}
with $r \to 0$ the conformal boundary and $r\to\infty$ the horizon.
The charged scalar $\Phi$ satisfies the Euclidean equation
\begin{equation}
    \left( -\nabla^2 + m_{\rm eff}^2 + q^2 A_\mu A^\mu \right)\Phi = 0,
\end{equation}
where the near-horizon gauge field is (Euclidean continuation)
\begin{equation}
    A_{t_E}(r) = -\, \frac{i q E \ell^2}{r}.
\end{equation}
We define the Whittaker parameters
\begin{equation}
    \kappa = - i\, \frac{\ell^2 q E}{\sqrt{2}}, 
    \qquad 
    k = i\sqrt{2}\,\ell^2\omega,
\end{equation}
and the effective $\mathrm{AdS}_2$ scaling exponent
\begin{equation}
    \mu^2 = \frac{1}{4} + m_{\rm eff}^2 \ell^2 - \frac{\ell^4 q^2 E^2}{4}.
\end{equation}
The BF bound is violated when $\mu^2 < 0$.
\subsection{IR Infalling Boundary Condition}

The general radial solution in $x$-coordinates is
\begin{equation}
    R(x)= C_1\, M_{\kappa,\mu}\!\left(\frac{k}{x}\right)
        + C_2\, W_{\kappa,\mu}\!\left(\frac{k}{x}\right).
\end{equation}
Near the horizon ($x\to 0$, equivalently $r\to\infty$), the Whittaker functions behave as
\begin{equation}
    W_{\kappa,\mu}\!\left(\frac{k}{x}\right) 
    \sim \exp\!\left(-\frac{k}{2x}\right)\left(\frac{k}{x}\right)^{\kappa},
\end{equation}
\begin{equation}
    M_{\kappa,\mu}\!\left(\frac{k}{x}\right)
    \sim 
    \frac{\Gamma(1+2\mu)}{\Gamma(\frac12+\mu-\kappa)}
    \exp\!\left(+\frac{k}{2x}\right)\!\left(\frac{k}{x}\right)^{-\kappa}
    + \cdots.
\end{equation}
For $k = i\sqrt{2}\,\ell^2\omega$, the phases $\exp(\pm k/2x)$ are oscillatory.  
The \emph{infalling} condition selects the mode proportional to
$\exp(-ik/2x)$, hence
\begin{equation}
    R_{\rm IR}(x) \propto 
    W_{\kappa,\mu}\!\left(\frac{k}{x}\right).
\end{equation}
Thus we set $C_1=0$ and keep only the $W$-branch.

\subsection{UV Asymptotics and Extraction of \texorpdfstring{$A_{\text{±}}$}{A±}}
At small argument $z \to 0$, the Whittaker function expands as
\begin{equation}
\begin{split}
    W_{\kappa,\mu}(z)
    = 
    \frac{\Gamma(2\mu)}{\Gamma\!\left(\frac12+\mu-\kappa\right)}
    z^{\frac12-\mu}
    +
    \frac{\Gamma(-2\mu)}{\Gamma\!\left(\frac12-\mu-\kappa\right)}
    z^{\frac12+\mu}
    +\cdots.
\end{split}
\end{equation}
Setting $z=k/x$ and using $r = 1/x$ gives the near-boundary form
\begin{equation}
    R(r) = A_-(\omega)\, r^{\Delta_-} + A_+(\omega)\, r^{\Delta_+},
\qquad
    \Delta_\pm = \frac12 \pm \mu,
\end{equation}
with coefficients
\begin{equation}
\begin{aligned}
A_-(\omega)
&=
C_2\, k^{\frac12-\mu}\,
\frac{\Gamma(2\mu)}{\Gamma\!\left(\frac12+\mu-\kappa\right)}, \\[4pt]
A_+(\omega)
&=
C_2\, k^{\frac12+\mu}\,
\frac{\Gamma(-2\mu)}{\Gamma\!\left(\frac12-\mu-\kappa\right)}.
\end{aligned}
\end{equation}
The normalization constant $C_2$ cancels in the ratio:
\begin{equation}\label{AminusAplusRatio}
\frac{A_-(\omega)}{A_+(\omega)} 
=
k^{-2\mu}\,
\frac{\Gamma(2\mu)}{\Gamma(-2\mu)}\,
\frac{\Gamma\!\left(\frac12-\mu-\kappa\right)}
     {\Gamma\!\left(\frac12+\mu-\kappa\right)}.
\end{equation}
\subsection{ Boundary Generating Functional}

The Euclidean on-shell action from the bulk plus counterterm 
$S_{\rm bdy}^{(1)}$ reduces to
\begin{equation}
    S_{\rm on-shell}
    =
    \int d\omega \,
    \left[ A_+^*(\omega)\, A_-(\omega)
         + A_-^*(\omega)\, A_+(\omega) \right],
\end{equation}
once we impose the alternate quantization condition $A_+=J(\omega)$, 
where $J(\omega)$ is the boundary source.\newline
Varying with respect to $J$ gives the one-point function 
$\langle A_-(\omega)\rangle$, and the connected two-point function satisfies
\begin{equation}
    \langle A_-(\omega) A_-^*(\omega')\rangle
    =
    2\pi\delta(\omega-\omega')\, G^{(0)}(\omega),
\end{equation}
with
\begin{equation}\label{BarePropagatorDef}
G^{(0)}(\omega)
= 
-\,\left[
\frac{A_-(\omega)}{A_+(\omega)}
+
\frac{A_-^*(\omega)}{A_+^*(\omega)}
\right].
\end{equation}
Using \eqref{AminusAplusRatio}, one obtains the explicit form
\begin{equation}
G^{(0)}(\omega)
=
k^{-2\mu}
\frac{4\,\Gamma(2\mu)}{\Gamma(1-2\mu)}
\frac{\Gamma\!\left(\frac12-\mu-\kappa\right)}
     {\Gamma\!\left(\frac12+\mu-\kappa\right)}.
\end{equation}
This is the analogue of eq.~(A.13) in \cite{Aharony:2023amq,faulkner2011emergent,iqbal2010quantum,Iqbal:2011aj}.

\bibliographystyle{apsrev4-1}
\bibliography{biblio}

\end{document}